\documentclass[reprint,amsmath,amssymb,aps, physrev,nofootinbib]{revtex4-2}

\usepackage{graphicx}
\usepackage{dcolumn}
\usepackage{bm}
\usepackage{xspace}
\usepackage{float}
\usepackage{mathrsfs}
\usepackage{dsfont}
\usepackage{physics}
\usepackage{pgfplots}
\pgfplotsset{compat=1.16}
\usepackage[colorlinks=true,linkcolor=blue,urlcolor=blue,citecolor=blue]{hyperref}

\newcommand{\filterfunctions}{\texttt{filter\-\_functions}\xspace}
\newcommand{\ts}[1]{\textsuperscript{#1}}
\newcommand{\mc}[1]{\ensuremath{\mathcal{#1}}}
\newcommand{\mr}[1]{\ensuremath{\mathrm{#1}}}
\newcommand{\im}[0]{\ensuremath{\mr{i}\xspace}}
\newcommand{\e}[0]{\ensuremath{\mr{e}\xspace}}
\newcommand{\ad}{\ensuremath{^\dagger}\xspace}
\newcommand{\gth}[1]{\ensuremath{^{(#1)}}\xspace}
\newcommand{\ctrlmat}[0]{\ensuremath{\mc{\Tilde{B}}}\xspace}
\newcommand{\liouvilleL}[0]{\ensuremath{\mc{Q}}\xspace}
\newcommand{\entinfid}[0]{\ensuremath{\mc{I}_{\alpha}}\xspace}
\newcommand{\uh}[1]{\ensuremath{u_h(#1)}}
\newcommand{\dR}[4]{\ensuremath{\pdv{\ctrlmat_{#1}^{#2}{}(#3)}{\uh{#4}}}}
\newcommand{\dL}[3]{\ensuremath{\pdv{\mc{Q}_{#1}\gth{#2}{}}{\uh{#3}}}}

\begin{document}
\preprint{APS/123-QED}

%TC:ignore
\title{Analytic Filter Function Derivatives for Quantum Optimal Control}

%--------------------------
%Authors
%--------------------------
\author{Isabel Nha Minh Le}
\email{isabel.le@rwth-aachen.de}
\author{Julian D. Teske}
\author{Tobias Hangleiter}
\author{Pascal Cerfontaine}
\author{Hendrik Bluhm}
\email{bluhm@physik.rwth-aachen.de}
\affiliation{
JARA-FIT Institute for Quantum Information, Forschungszentrum Jülich GmbH and RWTH Aachen University,
52074 Aachen, Germany
}

\date{\today}

%%%%%%%%%%%%%%%%%%%%%%%%%%%%%%%%%%%%%%%%%%%%%%
%Abstract
%%%%%%%%%%%%%%%%%%%%%%%%%%%%%%%%%%%%%%%%%%%%%%
\begin{abstract}
Auto-correlated noise appears in many solid state qubit systems and hence needs to be taken into account when developing gate operations for quantum information processing. However, explicitly simulating this kind of noise is often less efficient than approximate methods. Here, we focus on the filter function formalism, which allows the computation of gate fidelities in the presence of auto-correlated classical noise. Hence, this formalism can be combined with optimal control algorithms to design control pulses, which optimally implement quantum gates. To enable the use of gradient-based algorithms with fast convergence, we present analytically derived filter function gradients with respect to control pulse amplitudes, and analyze the computational complexity of our results. When comparing pulse optimization using our derivatives to a gradient-free approach, we find that the gradient-based method is roughly two orders of magnitude faster for our test cases. We also provide a modular computational implementation compatible with quantum optimal control packages. 
\end{abstract}
%TC:endignore

\maketitle

%%%%%%%%%%%%%%%%%%%%%%%%%%%%%%%%%%%%%%%%%%%%%%
%Main Paper
%%%%%%%%%%%%%%%%%%%%%%%%%%%%%%%%%%%%%%%%%%%%%%

%---------------------------------------------
\section{\label{sec:Introduction}Introduction}
%---------------------------------------------
Noise leading to the loss of quantum information remains a major challenge in the current development of quantum computers \cite{Nielsen2010c, Harrow2017, Unruh1995, Preskill2018}. Minimizing decoherence, while still leaving the system accessible for quantum control, stands in the center of quantum optimal control \cite{Glaser2015, Peirce1988, Rabitz2009}. 
While quasi-static noise can be well addressed in pulse optimization approaches, treating so-called colored noise characterized, e.g., by $1/f^\alpha$ spectral noise densities remains a difficult task \cite{Wang2012,Wang2014,Wang2014a,Yang2016}. However, colored noise poses a major noise contribution in many candidate systems for quantum information processing. As such, leading solid state qubit implementations are subject to colored flux or charge noise \cite{Wellstood1987,Bylander2011,Drung2011,Anton2012,Kuhlmann2013,Yoneda2018,Struck2020}.

The filter function formalism is a suitable and experimentally verified tool to fully describe a quantum system under wide-sense stationary classical noise with arbitrary auto- and cross-correlations \cite{Green2013,Cerfontaine2020,Hangleiter2020,Soare2014}. A so-called filter function quantifies the noise susceptibility of a quantum channel as a function of noise frequency. Specifically, the average gate infidelity \cite{Nielsen2002,Green2012}, a useful metric for the accuracy of quantum operations, is accessible via filter functions. Hence, this formalism can be used to design cost functions for the  optimization of control pulses in the presence of correlated noise. 

In previous works, colored noise has been taken into account by combining Monte Carlo simulations with Nelder-Mead optimization \cite{Huang2017, Huang2019}. In addition, filter function gradients have been used for quantum optimal control using gradient-based algorithms. Such algorithms typically require fewer iterations, but the overall performance gain depends on the cost of evaluating gradients. While for many problems, this cost turns out prohibitively large, it was shown that unitary quantum dynamics allow gradients to be computed rather efficiently \cite{Kuprov2009}. Filter function gradients  were calculated either by auto-differentiation \cite{Ball2020} or finite differences \cite{Cerfontaine2014}. Here, we develop analytical filter function gradients, study their computational complexity, and benchmark their application for pulse optimization. This not only allows for a numerically robust and computationally efficient implementation, but also provides insight if and when efficiency gains are expected compared to other methods. Additionally, no specialized software packages, e.g. for automatic differentiation, are required.

This paper is organized as follows: In Section~\ref{sec:Filter Function Formalism}, we introduce the theoretical concepts of the filter function formalism for an easily accessible, but still non-trivial case of a single pulse of one control operator and one noise source. Subsequently, we show how to obtain the derivatives of the filter function for this case in Section~\ref{sec:Filter Function Derivatives} (a full but more complex derivation is presented in the appendix). We continue to introduce the numerical implementation of the derivatives in Section~\ref{sec:Implementation}, and present an analysis of its computational complexity in Section~\ref{sec:Computational Complexity}. In Section~\ref{sec:Application}, we apply the newly implemented derivatives to pulse optimization and compare the results with a gradient-free optimization approach. We conclude in Section~\ref{sec:Conclusion} by summarizing our results and giving an outlook.

For the purpose of conciseness, we use the following notation: We denote operators and their matrix representations by Roman font, e.g. $P$, and reserve calligraphic font for quantum operations and their representations, e.g. $\mc{P}$. Additionally, we denote the control matrix, which we will introduce in the following section, with $\ctrlmat$, due to its resemblance with the Liouville representation of quantum operations. Operators in the interaction picture are written with an overset tilde, e.g. $\Tilde{P}=U^\dag P U$ with the toggling-frame operator $U$. Furthermore, we use an overset bar for the matrix representation of operators transformed into the eigenbasis of a Hamiltonian, e.g. $\Bar{P}=V^\dag P V$ with the corresponding unitary matrix of eigenvectors $V$. A general matrix is denoted by DS font, e.g. $\mathds{X}$, while its element-wise notation is labeled as $\left[ \mathds{X} \right]_{pq}$. Lastly, we denote the identity matrix by $\mathds{1}$, and set $\hbar\equiv 1$ throughout this work. 

%-----------------------------------------------------------------------
\section{\label{sec:Filter Function Formalism}Filter Function Formalism for a single pulse}
%-----------------------------------------------------------------------
Before computing the derivatives, we first review the filter function formalism \cite{Green2012,Green2013,Cerfontaine2020,Hangleiter2020} for the simple, but non-trivial case of a single pulse with one control operator with variable amplitude and one noise source. To this end, we break down the quantum system's Hamiltonian into control and noise contributions, and derive the ensemble average entanglement infidelity as well as the filter function. The latter is a central quantity of the formalism and describes the quantum gate's susceptibility to noise as a function of noise frequency. Since we focus on the infidelity, we summarize the derivation given by Green et al. \cite{Green2012}, even though more general approaches exist \cite{Cerfontaine2020,Hangleiter2020}.

We start by considering a quantum system, whose total Hamiltonian $H(t)$ during $t\in [0,\tau]$ consists of two parts: a control Hamiltonian $H_c(t)$ consisting of a time-independent contribution $H_0$ and adjustable parameters to achieve the intended quantum operation, and a noise Hamiltonian $H_n(t)$ perturbing $H_c(t)$, 
\begin{subequations}\label{eq:Hamiltonian}
\begin{align}
    H(t) &= H_c(t) + H_n(t)
    \\
    &= \left[ H_0 + u(t)A \right] + b(t)B.
\end{align}
\end{subequations}
Here, $u(t)$ is a single time-dependent control amplitude with the corresponding control operator $A$. Noise enters via the random variable $b(t)$ and the corresponding Hermitian noise operator $B$. 
The time evolution generated by the total Hamiltonian is described by the unitary operator $U(t) = \exp\left( -\im \int_0^t H(t^\prime) dt^\prime \right)$. 
It is possible to rewrite the total propagator by factoring it into two parts as $U(t) = U_c(t) \Tilde{U}(t)$, where $U_c(t)$ contains the control effects, and the unitary operator $\Tilde{U}(t)$ captures the effect of a single noise realization. 
While the control propagator $U_c(t)$ fulfills the noise-free Schrödinger equation $\im \pdv{U_c(t)}{t} = H_c(t)U_c(t)$, it can be shown \cite{Haeberlen1968} that $\Tilde{U}(t)$ fulfills the equation of motion $\im \pdv{\Tilde{U}(t)}{t} = \Tilde{H}_n(t) \Tilde{U}(t)$, with $\Tilde{H}_n(t) = U_c^\dag(t) H_n(t) U_c(t)$ the noise Hamiltonian transformed into the interaction picture of the control Hamiltonian. 
The generator of $\Tilde{U}(t=\tau) \equiv \Tilde{U}$ can then be considered as a time-independent effective Hamiltonian $H_{\text{eff}}$, such that 
\begin{equation}
    \Tilde{U} = \exp\left( -\im H_{\text{eff}}\cdot\tau \right). \label{eq:noise propagator}
\end{equation}
Using the Pauli basis, $H_\text{eff}$ can be written in terms of an error vector $\vec{\beta}$ as $H_\text{eff} = \vec{\beta}\cdot \vec{\sigma}$. 
Eq.~\eqref{eq:noise propagator} can be expanded by using the Magnus expansion \cite{Blanes2009b,Magnus1954}, such that the exponent is given by $H_\text{eff} = \sum_{\mu=1}^\infty \vec{\beta}_\mu \cdot \vec{\sigma}$ with the first order term
\begin{equation}
   H_{\text{eff},1} = \vec{\beta}_1 \cdot \vec{\sigma} = \frac{1}{\tau} \int_0^\tau dt \Tilde{H}_n(t). \label{eq:ME1 beta} 
\end{equation}
For small noise strength, i.e. sufficiently small $\vec{\beta}_i$, higher orders provide diminishing contributions \cite{Green2012,Hangleiter2020}.

A suitable measure for the quantum gate accuracy is the ensemble average entanglement infidelity \cite{Nielsen2002}, to which we will refer to as simply the infidelity $\mc{I}$. The infidelity is linked to $\vec{\beta}$ via
\begin{equation}
    \mc{I} \equiv \langle \mc{I}(\tau) \rangle = \frac{1}{2} \left[ 1 - \langle \cos\left( 2|\vec{\beta}| \right) \rangle \right]. \label{eq:entinfid1}
\end{equation}

By Taylor expanding the cosine term in Eq.~\eqref{eq:entinfid1}, the infidelity can be approximated by $\mc{I} = \frac{1}{2}\langle |\vec{\beta}|^2\rangle$ for small noise $|\vec{\beta}|\ll 1$. Evaluating this term only requires the square of the error vector $|\vec{\beta}|^2 = \left( \sum_k \beta_k^2 \right)$.
Inserting the first order term of the Magnus expansion given by Eq.~\eqref{eq:ME1 beta} leads to 
\begin{equation}
    \langle \beta_{1,k}^2 \rangle = \int_0^\tau dt_1 \int_0^\tau dt_2 \left( \langle b(t_1)b(t_2) \rangle \ctrlmat_k(t_1) \ctrlmat_k(t_2) \right), \label{eq:beta-squared}
\end{equation}
where $\langle b(t_1)b(t_2)\rangle$ is the noise amplitude's auto-correlation function, and $\ctrlmat_k(t)$ denotes the control matrix elements in the time domain defined by
\begin{equation}
    \ctrlmat_k(t) = \tr\left( U_c^\dag(t) B U_c(t) \sigma_k \right), \label{eq:ctrlmat_t}
\end{equation}
where we now consider the matrix representation of the noise operator $B$.
Note that in our case of a single noise contribution, the control matrix reduces to a control vector.

Under the assumption of wide-sense stationary classical noise, the spectral noise density $S(\omega)$ can be defined as the Fourier transform of $\langle b(t_1)b(t_2)\rangle$. Assuming  the noise to be Gaussian, which is reasonable in many cases \cite{Szakowski2017}, $S(\omega)$ gives a full characterization of the noise. Inserting $S(\omega)$ into Eq.~\eqref{eq:beta-squared} and shifting the Fourier transformation to the control matrix in frequency domain defined by
\begin{equation}
    \ctrlmat(\omega) = \int_0^\tau dt \ctrlmat(t) \e^{\im\omega t}, \label{eq:ctrlmat_omega}
\end{equation}
results in $\langle \beta_{1,k}^2 \rangle = \int \frac{d\omega}{2\pi} \ctrlmat_k^\ast(\omega) S(\omega) \ctrlmat_k(\omega)$. By summing over the Cartesian coordinates, we obtain
\begin{subequations}\label{eq:entinfid2}
\begin{align}
    \mc{I} &= \frac{1}{2}\int_{-\infty}^\infty \frac{d\omega}{2\pi} \left( \sum_k |\ctrlmat_k(\omega)|^2 \right) S(\omega)\\
    &= \frac{1}{2}\int_{-\infty}^\infty \frac{d\omega}{2\pi} F(\omega)S(\omega), 
\end{align}
\end{subequations}
with the filter function
\begin{equation}
    F(\omega) = \sum_k |\ctrlmat_k(\omega)|^2. \label{eq:FF}
\end{equation}
This expression gives a full description of how the noise contribution affects the given quantum channel described by $\ctrlmat(\omega)$.

%-----------------------------------------------------------------------
\section{\label{sec:Filter Function Derivatives}Filter Function Derivatives for a single pulse}
%-----------------------------------------------------------------------
In the following, the filter function gradient with respect to the control amplitude is derived for the previously presented simple case with only a single control variable and noise contribution. To this end, we approximate $u(t)$ for $t\in [0, \tau]$ to be a sequence of $n_{\Delta t}$ piecewise-constant control amplitudes $\{ u_1, u_2, ..., u_{\Delta t} \}$ and restrict our derivation to the case of a single constant control amplitude $u\in\mathds{R}$.

Differentiating Eq.~\eqref{eq:entinfid2} requires the filter function derivative with respect to the control amplitude $u$,
\begin{equation}
    \pdv{F(\omega)}{u} = 2 \Re\left(\sum_k  \ctrlmat_{k}^\ast (\omega) \pdv{\ctrlmat_{k}(\omega)}{u} \right). \label{eq:dF}
\end{equation}
By applying the product rule, the gradient of the control matrix in frequency space, $\pdv{\ctrlmat_{k}(\omega)}{u}$, is given by the Fourier transform of
\begin{align}
\frac{\partial {\cal \Tilde{R}}_{k}(t)}{\partial u} &= \mathrm{tr}\left( \frac{\partial U^\dagger_c(t)}{\partial u} B U_c(t) \sigma_k + U_c^\dagger(t) B \frac{\partial U_c(t)}{\partial u} \sigma_k\right). \label{eq:ctrl_deriv}
\end{align}

In order to deduce the gradient of the control propagator, $\pdv{U_c(t)}{u}$, we consider a small perturbation $\delta u$ on $u$ and add a corresponding term to the control Hamiltonian, $H_c + H_\delta = \left(u + \delta u\right)A$. 
Let $U_c(t)$ be the control propagator of the unperturbed control Hamiltonian $H_c$. The perturbation Hamiltonian in the interaction picture defined by $H_c$ is then given by $\Tilde{H}_\delta(t) = U_c^\dagger (t) \left(\delta u \cdot A\right) U_c(t)$. The corresponding Schrödinger equation 
\begin{equation}
    \im \pdv{\Tilde{U}_\delta(t)}{t} = \Tilde{H}_\delta (t) \Tilde{U}_\delta(t) \label{eq:SG}
\end{equation}
defines the perturbation propagator $\Tilde{U}_\delta(t)$. 
A solution of Eq.~\eqref{eq:SG} can be approximated as an exponential operator by using the Magnus expansion up to first order similarly to Eqs.~\eqref{eq:noise propagator} and \eqref{eq:ME1 beta}, such that
\begin{subequations}\label{eq:U_delta}
\begin{align}
    \Tilde{U}_\delta(t) &= \exp\left( -\im \left( \sum_{i=1}^\infty \Omega_i \right) \right), \label{eq:U_delta2}\\
    \Omega_1 &= \int_0^t dt^\prime \Tilde{H}_\delta(t^\prime). \label{eq:ME exponent}
\end{align}
\end{subequations}
We introduce the basis change matrix $V$ consisting of the eigenvectors of the matrix representation of the control Hamiltonian $H_c$, which transforms into the eigenbasis of the control Hamiltonian and indicate an operator in matrix representation $P$ under such a transformation as $\Bar{P}=V^\dag P V$. 
Since $VV^\dag = V^\dag V=\mathds{1}$, inserting the identity into the exponent given by Eq.~\eqref{eq:ME exponent} leads to
\begin{align}
    \Omega_1 &= V \left( \int_0^t dt^\prime \Bar{U}_c^\dag(t^\prime)\Bar{\Tilde{H}}_\delta \Bar{U}_c(t^\prime) \right) V^\dag = V \mathds{K}(t) V^\dag \label{eq:Omega1}
\end{align}

Naturally, the control propagator in the eigenbasis of the control Hamiltonian $\Bar{U}_c$ is diagonal. 
Specifically, if we consider the sorted set of eigenvalues of the control Hamiltonian $\{\omega_i\}$ in the same order as the eigenvectors in $V$, it is given by $\Bar{U}_c = \textrm{diag}\left( \e^{\im \omega_i t} \right)$.
For the further calculation, we introduce the element-wise notation $\left[ \mathds{X} \right]_{pq}$ for the matrix $\mathds{X}$. The integral $\mathds{K}$ in Eq.~\eqref{eq:Omega1} can then be evaluated as
\begin{subequations} \label{eq:K}
\begin{align}
    \left[\mathds{K}(t)\right]_{pq} &= \left[ \Bar{\Tilde{H}}_\delta \right]_{pq} \int_0^t dt^\prime \exp\left( \im \left( \omega_p - \omega_q \right) t^\prime \right) \\
    &= \left[ \Bar{\Tilde{H}}_\delta\right]_{pq} \left[ \delta_{pq} t + (1-\delta_{pq}) \frac{\mathrm{e}^{\im (\omega_p -\omega_q)t} - 1}{\im (\omega_p - \omega_q)} \right] \\
    &= \left[\Bar{\Tilde{H}}_\delta\right]_{pq} \left[\mathds{M}(t)\right]_{pq},
\end{align}
\end{subequations}
where $\delta_{pq}$ is the Kronecker delta.
We now notice that the total propagator of $H_c + H_\delta$ is given by $U(t) = U_c(t) \Tilde{U}_\delta(t)$, and that $\Tilde{U}_\delta(t)\mid_{\delta u=0}=\mathds{1}$. Using this, we can rewrite the control propagator derivative as 
\begin{align}\label{eq:relation dU}
    \pdv{U_c(t)}{u} &= \pdv{U(t)}{(\delta u)}\mid_{\delta u = 0}
\end{align}
In order to calculate $\pdv{U(t)}{(\delta u)}\mid_{\delta u = 0}$, we differentiate Eq.~\eqref{eq:U_delta2} to first order. 
To this end, we need to take the derivative of the exponent given by Eq.~\eqref{eq:Omega1}. 
Taking into account Eq.~\eqref{eq:K} and recognizing that
\begin{align}
    \pdv{\Bar{\Tilde{H}}_\delta}{(\delta u)} = \pdv{}{(\delta u)}\left( V^\dag \left( \delta u\cdot A \right) V^\dag \right) = \Bar{A},
\end{align}
the derivative of the exponent is given by
\begin{align}
    \pdv{\Omega_1}{(\delta u)} %&= V \left[\mathds{M}(t)\right]_{pq} \pdv{\left[\Bar{\Tilde{H}}_\delta\right]_{pq}}{(\delta u)}\mid_{\delta u = 0} V^\dag\\
    &= V \left(\mathds{M}(t) \circ \Bar{A}\right) V^\dag, \label{eq:dOmega}
\end{align}
where $(\circ)$ symbolizes element-wise multiplication.
Considering Eqs.~\eqref{eq:U_delta} - \eqref{eq:dOmega}, the derivative of the control propagator with respect to the constant control amplitude can be calculated as
\begin{align}\label{eq:dU}
    \pdv{U_c(t)}{u} = -\im U_c(t) V \left( \mathds{M}(t) \circ \Bar{A}\right) V^\dag.
\end{align}

By inserting Eq.~\eqref{eq:dU} into Eq.~\eqref{eq:ctrl_deriv} and the latter equation into Eq.~\eqref{eq:dF}, we have derived a complete analytic form of the filter function gradient for the considered case of a single pulse.

In general, a pulse sequence can consist of several time steps, control contributions, and noise sources. Under the assumption of piecewise-constant control amplitudes, the total control effect on the quantum system is a composition of the single pulse effects. The total control matrix is then given by \cite{Hangleiter2020, Cerfontaine2020}
\begin{align}
    \ctrlmat(\omega) &= \sum_{g=1}^{n_{\Delta t}} \e^{\im \omega t_{g-1}} \ctrlmat\gth{g}(\omega)\mathcal{Q}\gth{g-1}, \label{eq:R}
\end{align}
where $\ctrlmat\gth{g}(\omega)$ are the single pulse control matrices given by Eqs.~\eqref{eq:ctrlmat_t} and \eqref{eq:ctrlmat_omega}, and $\mathcal{Q}\gth{g-1}$ are the cumulative control propagators of the individual single pulses, i.e. the control propagators in Liouville representation multiplied with each other up to each time step. A more detailed explanation is given in Appendix~\ref{ap:Filter Function Formalism}. 
Consequently, when calculating filter function gradients for a general pulse sequence, products of $\ctrlmat\gth{g}(\omega)\mathcal{Q}\gth{g-1}$ in the total control matrix need to be taken into account in Eq.~\eqref{eq:dF}.
To this end, partial derivatives of the cumulative control propagators of the individual pulses have to be deduced additionally. A full calculation of the filter function gradient for a generic pulse sequence is given in Appendix~\ref{ap:FFD}.

%----------------------------------------------------------
\section{\label{sec:Implementation}Software Implementation}
%----------------------------------------------------------
We implemented the filter function gradients derived in the previous section as part of the open source software framework \filterfunctions \cite{Hangleiter2019a, Hangleiter2020}. This python package facilitates the efficient numerical calculation of generalized filter functions and derived quantities, such as the infidelity, for given pulse sequences. A pulse sequence is represented by the \texttt{PulseSequence} class, from which properties like the filter function or infidelity can directly be computed.
The newly implemented module \texttt{gradient} enables the calculation of filter function and infidelity derivatives for systems that are subject to classical wide-sense stationary noise, which can be characterized by an arbitrary spectral noise density. The gradients are taken with respect to stepwise constant control amplitudes. The function \texttt{infidelity\-\_derivative()} is placed at the user's disposal for direct calculation of the infidelity derivatives for a given pulse sequence. This function can be used directly for quantum optimal control, i.e. to optimize gate fidelities.
An illustration of the structure of the implementation can be found in Appendix~\ref{ap:AdditionalGraphics}, and a verification of the implemented analytical gradients are available in Ref.~\citenum{Hangleiter2020}. Furthermore, we ensured the implementation's compatibility with quantum optimal control packages, such as \texttt{qopt} \cite{qopt2020, Teske2020}.

%---------------------------------------------------------------------
\section{\label{sec:Computational Complexity}Computational Complexity}
%---------------------------------------------------------------------
In principle, our software implementation enables the calculation of filter function and infidelity gradients without any constraints on the quantum system's dimension $d$ or on the number of control and noise operators $n_c$ and $n_\alpha$. Furthermore, a pulse sequence can contain any number of time steps $n_{\Delta t}$ and the number of frequency samples $n_\omega$ describing the noise spectral density is completely variable.
Naturally, the computational complexity in computing the infidelity derivatives depends on the chosen set of parameters. Theoretical investigations of the implementation lead to the expected scaling behavior summarized in  Table~\ref{tab:complexity} for dominant terms.
\begin{table}[h]
\caption{\label{tab:complexity}Summary Computational Complexity}
\begin{ruledtabular}
\begin{tabular}{lcr}
\textrm{Parameter} & \textrm{Expectation} & \textrm{Run Time}\\
\colrule
Number of frequency samples $n_\omega$ & $\mathcal{O}(n_\omega)$ & $\mathcal{O}(n_\omega)$\\
Number of control operators $n_c$ & $\mathcal{O}(n_c)$ & $\mathcal{O}(n_c)$\\
Number of noise operators $n_\alpha$ & $\mathcal{O}(n_\alpha)$ & $\mathcal{O}(n_\alpha)$\\
Number of time steps $n_{\Delta t}$ & $\mathcal{O}(n_{\Delta t}^2)$ & $\mathcal{O}(n_{\Delta t}^{1.92})$ \\
Dimension $d$ & $\mathcal{O}(d^{b+4})$\footnote{$b$ arises from the multiplication of two $n\times n$-matrices, which scales polynomially with $n^b$. For a naive algorithm $b=3$ \cite{Cormen2009} and for the Coppersmith–Winograd algorithm $b=2.376$ \cite{Coppersmith1990}.} & $\mathcal{O}(d^{4.35})$\\
\end{tabular}
\end{ruledtabular}
\end{table}

We analyze the actual run time behavior by running the implemented software module with various pulses. To this end, we increase one of the parameters $n_\alpha$, $n_c$, $n_\omega$, $n_{\Delta t}$ and $d$, while fixing the remaining quantities and generating the control amplitudes randomly for each pulse. We then obtain the scaling behavior of the implementation by means of asymptotic fits. A graphical illustration can be seen in Fig.~\ref{fig:runtimepoly} and in Appendix~\ref{ap:AdditionalGraphics}. Table~\ref{tab:complexity} contrasts the actual run time results with their theoretical expectations.
\begin{figure}[H]
    \centering
    \includegraphics[width=\the\columnwidth]{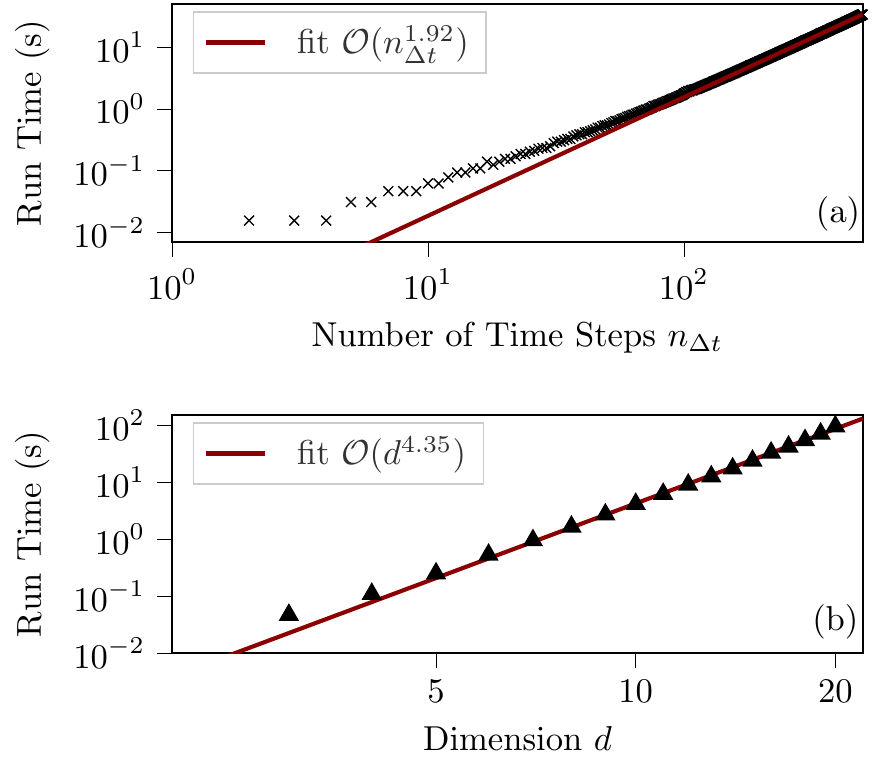}
    \caption{Run time behavior for computing the infidelity gradients on a standard desktop computer\footnote{We used a Intel\textregistered Core\textsuperscript{TM}i5-2400 processor with four logical cores.} for (a) various numbers of time steps $n_{\Delta t}$ and (b) various dimensions $d$. (a) shows a random pulse sequence with dimension $d=2$, $n_\omega = 200$ frequency samples, $n_c=n_\alpha=2$ control and noise operators, and $n_d=1$ drift direction. A fit to the data yields the predicted quadratic scaling behavior. (b) shows the median run time of 50 randomly generated pulse sequences of $n_{\Delta t}=3$ time steps. The parameters chosen for each random pulse sequence were: $n_\omega = 200$ frequency samples, $n_c=n_\alpha=2$ control and noise operators, and $n_d=1$ drift direction. A fit to the data yields a polynomial scaling behavior.}\label{fig:runtimepoly}
\end{figure}

The plots confirm the expected linear dependence on $n_\omega$, $n_c$ and $n_\alpha$.
For various $n_{\Delta t}$ and $d$, the run time data is shown in Fig.~\ref{fig:runtimepoly}.
A polynomial fit on the data for various $n_{\Delta t}$ stands in agreement with the expectation of quadratic scaling. Concerning the $d$-dependency, a clear polynomial behavior is visible. Due to memory limitation, we tested dimensions restricted to $d\leq 20$. Within this restriction, lower order terms in $d$ dominate the scaling behavior, such that this cannot be considered as an asymptotic regime. Therefore, the underestimated exponent for the $d$-dependency does not contradict the theoretical expectation.

%-------------------------------------------
\section{\label{sec:Application}Application}
%-------------------------------------------
As mentioned previously, filter function derivatives lend themselves to gradient-based pulse optimization. In the following, we motivate the use of such gradient-based methods by comparing the numerical optimization of control pulses using filter function gradients to a gradient-free approach.

To this end, we use the filter function derivatives in conjunction with \texttt{scipy}'s L-BFGS-B algorithm \cite{Zhu1997, Morales2011} and contrast this strategy with a gradient-free constrained Nelder-Mead method \cite{Nelder1965, constrNMPy} in terms of run time and error rates of the optimized pulses. We conduct both optimization approaches within the quantum optimal control package \texttt{qopt} \cite{qopt2020, Teske2020}. To facilitate a fair comparison, we disabled the internal multi-threading of the optimization algorithms in python.

We apply each technique to a generic 4-level quantum system corresponding to the optimization of two-qubit gates. For this purpose, we use a control Hamiltonian given by 
\begin{align}
  H_c = \sum_{ij} u_{ij,g}\cdot \sigma_i \otimes \sigma_j,
\end{align}
where $u_{ij,g}$ signifies the piecewise constant control amplitudes at the discrete time step $g~\in~(1,2\dots,n_{\Delta t})$ of uniform length $\Delta t$. 
For simplicity, we assume the system to be exposed to exactly one noise source, such that the noise Hamiltonian is given by $H_n(t) = b(t) \cdot\sigma_0 \otimes \sigma_x$. In the latter, $b(t)$ denotes a noise amplitude corresponding to a pink noise spectral density of the form $S(f) = S_0 / f$ with frequency $f$ and constant $S_0$.

We choose the cost function of the pulse optimization to be the total infidelity $\mc{I}$, which is the sum of systematic and noise-induced deviations from the target gate. The former ones are quantified by the standard entanglement infidelity and the latter ones are calculated as in Eq.~\eqref{eq:entinfid2}. The optimization problem then lies in the minimization of the infidelity $\mc{I}$ by finding the optimal control amplitudes $u_{ij,g}$.

\begin{figure}%[H]
     \centering
     \includegraphics[width=\the\columnwidth]{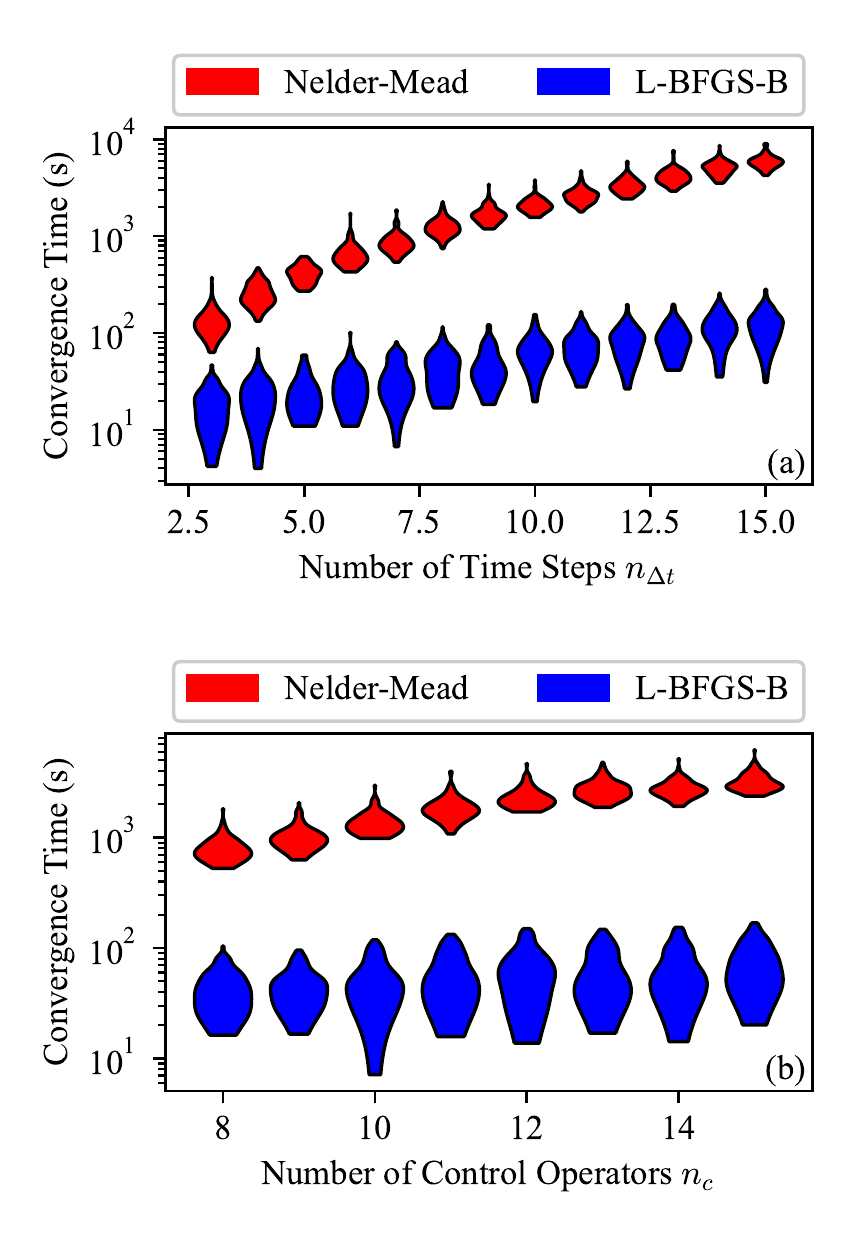}
     \caption{Convergence time of the analytical derivatives in conjunction with the L-BFGS-B algorithm compared to Nelder-Mead. The distribution of convergence times is depicted by the width of the violin plots, where each violin plot shows the distribution of 100 runs. The convergence time is (a) displayed as a function of the number of time steps, while the number of control operators is constant $n_c = 8$, and (b) as a function of number of control operators for a constant number of time steps $n_{\Delta t }=6$. In both cases the L-BFGS-B algorithm converges faster to optimal pulses with about the same infidelity as the Nelder Mead algorithm and scales better with an increasing number of degrees of freedom.}
      \label{fig:convergence_time}
\end{figure}

As convergence criterion, we choose that $\mc{I}$ improves by less than $10^{-7}$ within one iteration of the optimization algorithm, which does not favor either algorithm (see Fig.~\ref{fig:convergence_runs} in Appendix~\ref{sec:ComparisonNM}). The performance is depicted in Fig.~\ref{fig:convergence_time}, where we observe that the L-BFGS-B algorithm converges faster and scales better with the number of time steps $n_{\Delta t}$ and the number of control operators $n_c$. Both algorithms find locally optimal pulses with similar fidelities (see Fig.~\ref{fig:final_infid} in Appendix~\ref{sec:ComparisonNM}). 

In Appendix~\ref{ap:FFD}, we discuss the convergence of the two optimization algorithms in greater detail, demonstrating that the Nelder-Mead algorithm offers better global performance, while the L-BFGS-B algorithm is more prone to get stuck in local minima.
A further discussion of the benefits of filter functions compared to Monte Carlo methods and the description of open quantum systems by master equations can be found in Ref.~\citenum{Teske2020}.

%-----------------------------------------------------
\section{\label{sec:Conclusion}Conclusion and Outlook}
%-----------------------------------------------------
In this paper, we presented analytically derived filter function gradients and their numerical implementation. By doing so, we make the gradients easily accessible for various pulse optimization algorithms. Furthermore, we conducted an analysis of the computational complexity for obtaining the filter function derivatives by our implementation. We verified the theoretical prediction by comparing them to the actual run time scaling behavior. Finally, we applied our filter function derivatives to gradient-based pulse optimization of generic two-qubit gates and contrast this approach with a gradient-free optimization method. While both strategies result in optimized pulses of similar fidelities, we showed that the gradient-based optimization requires up to two orders of magnitude less time to converge than the gradient-free optimization for our test case.

In addition to pulse optimization for implementation of quantum gates, various other applications for analytic filter function derivatives exist. 
Since the filter function formalism can be used to describe quantum algorithms in terms of pulse sequences \cite{Cerfontaine2020, Hangleiter2020}, filter function gradients can not only be used to optimize quantum gates, but also the implementation of quantum algorithms.
Furthermore, pulse optimization can also aid noise spectroscopy \cite{Pozza2019}, where carefully designed control pulses are used to obtain an accurate description of the present spectral noise density.
As another application, filter function derivatives could be used to assess the impact of quasistatic noise or calibration errors on the high-frequency noise properties of a given pulse. 
Lastly, if the noise environment changes, e.g. due to dynamically changing operational parameters of on-chip control electronics, our gradients can be used to quickly re-calibrate qubits in a quantum processor.

Based on our results analytic filter function derivatives can facilitate quantum optimal control for various proposed qubit implementations that are subject to arbitrary classical auto- and cross-correlated noise. Thus, these derivatives can help to assess the potential performance of candidate hardware platforms for quantum information processing \cite{Kelly2014, Cerfontaine2020b, Biercuk2009}.

\begin{acknowledgments}
This work is supported by the European Research Council (ERC) under the European Union’s Horizon 2020 research and innovation program (Grant Agreement No. 679342).

All correspondence should be addressed to Hendrik Bluhm.
\end{acknowledgments}

%%%%%%%%%%%%%%%%%%%%%%%%%%%%%%%%%%%%%%%%%%%%%%
%Appendix
%%%%%%%%%%%%%%%%%%%%%%%%%%%%%%%%%%%%%%%%%%%%%%
%TC:ignore
\appendix

%-----------------------------------------------------------------------
\section{\label{ap:Filter Function Formalism}Filter function formalism for a general pulse sequence}
%-----------------------------------------------------------------------

In Section~\ref{sec:Filter Function Formalism} of the main paper, the filter function formalism has been presented for the special case of a single control contribution and a single noise source. The following section recaptures some modifications needed for the calculation of filter function derivatives of a general pulse sequence consisting of various control and noise contributions.

Consider now the general description of a quantum system, whose total Hamiltonian $H(t)$ during $t\in[0,\tau]$ consists of two parts: a control Hamiltonian $H_c(t)$ consisting of a time-independent contribution $H_0$ and multiple adjustable parameters to achieve the intended quantum operation, and a noise Hamiltonian $H_n(t)$ perturbing $H_c(t)$,
\begin{subequations}
\begin{align}
    H(t) &= H_c(t) + H_n(t), \label{ap:eq:Hamiltonian}\\
    H_c(t) &=  H_0 + \sum_k u_k(t)A_k, \\
    H_n(t) &= \sum_\alpha s_\alpha(t) b_\alpha(t)B_\alpha.
\end{align}
\end{subequations}
Within $H_c(t)$, $u_k(t)$ is the adjustable control strength of the control operator $A_k$ at time $t$.
Likewise in $H_n(t)$, $b_\alpha(t)$ is the randomly distributed amplitude of the Hermitian noise operator $B_\alpha$ and $s_\alpha(t)$ captures the system's sensitivity to the corresponding noise source and might be dependent on the control amplitude.
    
Following the approach given in Section~\ref{sec:Filter Function Formalism}, the control matrix can be derived. The control propagator $U_c(t)$ again contains the control effects and fulfills the noise-free Schrödinger equation $\im\pdv{U_c(t)}{t}=H_c(t)U_c(t)$. In the main paper, the set of Pauli operators $\{\mathds{1},\sigma_x,\sigma_y,\sigma_z\}$ was chosen as an operator basis. In general, the operators can be expressed in any orthonormal operator basis $\{ C_0,C_1,...,C_{d^2-1} \} \in \mathds{C}^{d\times d}$ with respect to the Hilbert-Schmidt product $\langle C_i,C_j \rangle := \tr(C_i\ad C_j) = \mathds{1}\delta_{ij}$. 
Taking this generalization into account, and considering the noise sensitivity, the control matrix in time domain $\ctrlmat(t)$ needs to be modified. It is then defined as
\begin{equation}
    \ctrlmat(t)_{\alpha k} = s_\alpha(t) \cdot \tr \left( U_c\ad(t) B_\alpha U_c(t) C_k \right). \label{ap:eq:ctrlmat_t}
\end{equation}

In case of piecewise constant control, a control sequence consists of $n_{\Delta t}$ constant single pulses. Let us denote $[t_{g-1}, t_g]$ as the time interval corresponding to the $g$\ts{th} single pulse for $g\in\{1,...,n_{\Delta t}\}$. The control sequence can then be described with help of the single pulse propagators $U_c(t_g,t_{g-1})$. Consequently, the propagator cumulated up to the $g$\ts{th} time step is given by $Q_g = \prod_{l=g}^1 U_c(t_l, t_{l-1})$ and its so-called Liouville representation is defined by $\mc{Q}_{jk}\gth{g}{} = \tr\left( Q_g^\dag C_j Q_g C_k \right)$. Next, we denote the duration of the $g$\ts{th} single pulse with $\Delta t_g = t_g - t_{g-1}$, and the single pulse control matrix in frequency space at time step $g$ with 
\begin{equation}
    \ctrlmat\gth{g}(\omega)=\int_0^{\Delta t_g} dt \e^{\im \omega t}\ctrlmat\gth{g}(t). \label{ap:eq:R_g_omega}
\end{equation} The total control matrix of the pulse sequence can then be directly determined by
\begin{align}
    \ctrlmat(\omega) &= \sum_{g=1}^{n_{\Delta t}} \e^{\im \omega t_{g-1}} \ctrlmat\gth{g}(\omega)\mathcal{Q}\gth{g-1}. \label{ap:eq:R}
\end{align}
In the above, $\mathcal{Q}\gth{g-1}$ is the control propagator of the individual single pulses cumulated up to time step $g-1$. 
The temporal positions of the single pulses enter the expression due to the Fourier transform via the phase factor $\e^{\im \omega t_{g-1}}$. 
Therefore, it is possible to obtain the total control matrix of a generic pulse sequence by summing up each single pulse control matrix multiplied with the cumulated control propagators of the priorly executed single pulses.
Using Eq.~\eqref{ap:eq:R}, the filter function $F_\alpha$ for a noise contribution $\alpha$ can be obtained by Eq.~\eqref{eq:FF}. If we extend the calculation to generic dimensions $d$, the infidelity $\entinfid$ for a noise contribution $\alpha$ is given by
\begin{equation}
    \entinfid = \frac{1}{d} \int_{-\infty}^\infty \frac{d\omega}{2\pi} F_\alpha(\omega)S_\alpha(\omega).
\end{equation}

%----------------------------------------------------------------
\section{\label{ap:FFD}Filter function derivatives for a general pulse sequence}
%----------------------------------------------------------------
When considering a pulse sequence with $n_{\Delta t}>1$ time steps, the correlation terms in the total control matrix given by Eq.~\eqref{eq:R} need to be taken into account. In the following, we derive the filter function gradient for the general case of a pulse sequence under the assumption of piecewise-constant control.

We write $u_h(t_{g^\prime})$ for the control amplitude in direction $h$ at a fixed time $t_{g^\prime}$ and note that $u_h(t)=u_h(t_g)$ for $t\in[t_{g-1},t_g]$.
Applying the product rule on Eq.~\eqref{ap:eq:R} results in
\begin{align}\label{ap:eq:dR}
    \dR{\alpha k}{}{\omega}{t_{g^\prime}} &= \sum_{g=1}^{n_{\Delta t}} \sum_{j=1}^{d^2} \e^{\im\omega t_{g-1}} \cdot \\\nonumber
    &\cdot \left(  \dR{\alpha j}{(g)}{\omega}{t_{g^\prime}} \liouvilleL_{jk}\gth{g-1}{} + \ctrlmat_{\alpha j}\gth{g}{}(\omega) \dL{jk}{g-1}{t_{g^\prime}} \right).
\end{align}
This expression depends on four quantities: the control propagators in Liouville representation $\liouvilleL_{jk}\gth{g-1}{}$, and their partial derivatives $\dL{jk}{g-1}{t_{g^\prime}}$; and the control matrices $\ctrlmat_{\alpha j}\gth{g}{}(\omega)$, and their partial derivatives $\dR{\alpha j}{(g)}{\omega}{t_{g^\prime}}$. 
To break the calculation into comprehensive parts, we dedicate each of the mentioned quantities one of the following subsections.

%-------------------------------------------------------------
\subsection{\label{ap:subsec:Control matrix}Control matrix at time step $g$}
%-------------------------------------------------------------
For a fixed time step $g$, the single control matrix $\ctrlmat\gth{g}{}(\omega)$ is defined in Eq.~\eqref{ap:eq:R_g_omega}. In the following, we will evaluate the integral formula analytically. To this end, we introduce the basis change matrix $V\gth{g}{}$ with its columns being the eigenvectors of the control Hamiltonian at time step $g$, $H_c\gth{g}{}$. An operator $P$ in matrix representation transformed into the eigenbasis of $H_c\gth{g}{}$ is then denoted by $\Bar{P}\gth{g}{}=V\gth{g}{}^\dagger P V\gth{g}{}$.

Let $\{\omega_i\gth{g}{}\}_{1 \leq i \leq d}$ be the set of eigenvalues of $H_c\gth{g}{}$ in the same order as the eigenvectors in $V\gth{g}{}$. The control propagator during time step $g$ transformed into the eigenbasis of $H_c\gth{g}{}$ is naturally a diagonal matrix 
\begin{equation}
    D_{ij}\gth{g}{}(t) \equiv \left[\Bar{U}_c\gth{g}{}(t,t_{g-1})\right]_{ij} = \delta_{ij}\exp(-i\omega_i\gth{g}{}(t-t_{g-1})), \label{ap:eq:U_bar}
\end{equation}
with $t\in [t_{g-1},t_g]$.
Due to piecewise constant control, the sensitivity during the $g$\ts{th} time step is constant, i.e. $s_\alpha(t)=s_\alpha\gth{g}{}$. Inserting the identity $\mathds{1}=V\gth{g}{}^\dagger V\gth{g}{} = V\gth{g}{}V\gth{g}{}^\dagger$ into Eq.~\eqref{ap:eq:ctrlmat_t}, and taking into account the cyclic properties of the trace results in
\begin{subequations}
\begin{align}
    \ctrlmat\gth{g}{}(t)_{\alpha j} &= s_\alpha\gth{g}{} \tr\left( \Bar{U}_c\gth{g}{}^\dag \Bar{B}_\alpha \Bar{U}_c\gth{g}{} \Bar{C}_j \right) \\
    &= s_\alpha\gth{g}{} \tr\left( \sum_{ijkl} D_{ij}\gth{g}{}^\dag(t) \Bar{B}_{\alpha,jk} D_{kl}\gth{g}{}(t) \Bar{C}_{j,lm} \right)\\
    &= s_\alpha\gth{g}{} \sum_{ik} \left( \exp(i (\omega_i\gth{g}{} - \omega_k\gth{g}{}) t) \Bar{B}_{\alpha, ik} \Bar{C}_{j,ki} \right).
\end{align}
\end{subequations}
Now we can carry out the transformation of $\ctrlmat\gth{g}{}(t)$ into frequency domain given by Eq.~\eqref{ap:eq:R_g_omega}. Since $B_\alpha$ and $C_j$ are not time-dependent, evaluating the time integral over time dependent factors at time step $g$ yields
\begin{subequations}\label{ap:eq:integralI}
\begin{align} % matrix as mathds: I --> mathds{O}
    \mathds{O}_{ik}\gth{g}{}(\omega) &= \int_0^{\Delta t} dt \exp\left( \im \left( \omega - \omega_i\gth{g}{} - \omega_k\gth{g}{} \right) t \right) \\
    &= \frac{\exp\left( \im(\omega + \omega_i\gth{g}{} - \omega_k\gth{g}{})\Delta t_g \right) - 1}{\im(\omega + \omega_i\gth{g}{} - \omega_k\gth{g}{})}.
\end{align}
\end{subequations}
By inserting this expression into Eq.~\eqref{ap:eq:R_g_omega} and denoting element-wise multiplication by $(\circ)$, the control matrix in frequency space can be evaluated to
\begin{subequations} \label{ap:eq:R_g}
\begin{align}
    \ctrlmat_{\alpha j}\gth{g}{}(\omega) &= s_\alpha\gth{g}{} \sum_{ik} \Bar{B}_{\alpha,ik} \Bar{C}_{j,ki} \mathds{O}_{ik}(\omega) \\
    &= s_\alpha\gth{g}{} \tr \left( (\Bar{B}_{\alpha} \circ \mathds{O}\gth{g}{}(\omega)) \cdot  \Bar{C}_{j}) \right).
\end{align}
\end{subequations}

%------------------------------------------------
\subsection{\label{ap:subsec:Derivative control prop}Derivative of the control propagator}
%------------------------------------------------
In the main paper we have already derived a closed formula of the control propagator gradient in the case of a single pulse. The resulting expression in Eq.~\eqref{eq:dU} is valid for each control propagator gradient within a certain time step $g$. More specifically, the derivative of the control propagator $U_c(t,t_{g-1})$ within time step $g$ is given as
\begin{align}
    \frac{\partial U_c(t, t_{g-1})}{\partial u_h(t_g)} = -\im U_c(t, t_{g-1}) \cdot V\gth{g}{} \left( \mathds{M}\gth{g}{}(t) \circ \Bar{A}_h \right) \cdot V\gth{g}{}^\dag. \label{ap:eq:deriv_U_g}
\end{align}
The difference is that in the general case, various control contributions $h$ are considered, and that $V\gth{g}{}$ and $\mathds{M}\gth{g}{}(t)$ denote each quantity for the specific considered time step $g$.

%------------------------------------------------
\subsection{\label{ap:subsec:Derivative Liouville}Derivative of the propagator in Liouville representation}
%------------------------------------------------
The main difference between the cases of a single pulse and a pulse sequence is that we need to take the derivative of the cumulated control propagator in Liouville representation $\pdv{\liouvilleL_{jk}\gth{g-1}{}}{u_h(t_{g^\prime})}$ at each time step into account.
To this end, we keep in mind that $Q_g = \prod_{l=g}^1 U_c(t_l,t_{l-1}) = U_c(t_g,0)$ is the cumulative propagator up to time step $g$ and that its Liouville representation is given element-wise as $\mc{Q}_{jk}\gth{g}{} = \tr\left( Q_g^\dag C_j Q_g C_k \right)$.
For the further calculation we first need the derivative of $Q_g$, which can be evaluated as
\begin{widetext}
\begin{subequations}
\begin{equation}
    \pdv{Q_g}{u_h(t_{g^\prime})} = \pdv{U_c(t_{g-1},0)}{u_h(t_{g^\prime})} = \Theta(g-1, g^\prime) U_c(t_{g-1}, t_{g^\prime}) \cdot \pdv{U_c(t_{g^\prime}, t_{g^\prime -1})}{u_h(t_{g^\prime})} \cdot U_c(t_{g^\prime -1}, 0), \label{ap:eq:dU}
\end{equation}
\begin{equation}
    \Theta(g-1, g^\prime) =      
    \begin{cases}
        1 & g^\prime < g-1 \\
        0 & \text{otherwise}, \label{ap:eq:cases}
    \end{cases} 
\end{equation}
\end{subequations}
\end{widetext}
where Eq.~\eqref{ap:eq:cases} incorporates the fact that cumulative propagators up to a time step $g$ are independent of the control at a later point in time $g^\prime > g$ and where $\pdv{U_c(t_{g^\prime}, t_{g^\prime -1})}{u_h(t_{g^\prime})}$ given by Eq.~\eqref{ap:eq:deriv_U_g}.
By applying the product rule, we calculate the derivative of $\mc{Q}\gth{g-1}{}$ element-wise to 
\begin{widetext}
\begin{equation}
   \pdv{\mc{Q}_{jk}\gth{g-1}{}}{u_h(t_{g^\prime})} = \Theta(g-1, g^\prime) \tr\left( \pdv{Q_{g-1}^\dag}{u_h(t_{g^\prime})} C_j Q_{g-1} C_k +  Q_{g-1}^\dag C_j \pdv{Q_{g-1}{}}{u_h(t_{g^\prime})} C_k \right). \label{ap:eq:dL}
\end{equation}
\end{widetext}

%--------------------------------------------
\subsection{\label{ap:subsec:dR_g}Derivative of the control matrix} 
%--------------------------------------------
The last quantity that remains to be computed are the derivatives of single control matrices. In order to calculate the derivative of the control matrix at time step $g$ in the frequency domain $\pdv{\ctrlmat\gth{g}{}(\omega)}{u_h(t_{g^\prime})}$, we will first obtain a formula for the gradient in time space and subsequently transform the result into frequency space. To this end, we consider $t\in [t_{g-1}, t_g]$ and again label $\Delta t = t - t_{g - 1}$. If $s_\alpha(t)$ depends on the control amplitudes, an additive term including the noise sensitivity derivative needs to be taken into account. Here, we assume to know the analytic dependency of $s_\alpha(t)$ on the control amplitudes and therefore, concentrate on the remaining term by choosing $s_\alpha(t)$ to be independent of the control amplitudes. Using the product rule and cyclic properties of the trace, and keeping in mind that the noise and control operators are independent of the control strength leads to 
\begin{widetext}
\begin{align}
    \pdv{\ctrlmat_{aj}\gth{g}{}(\Delta t)}{u_h (t_{g^\prime})} &= \im \delta_{gg^\prime} s_\alpha\gth{g}{} \tr \bigg( U_c^\dag(t, t_{g^\prime-1}) B_\alpha U_c(t,t_{g^\prime-1}) \Big[C_j, V\gth{g}{}(\mathds{M}\gth{g}{}(t) \circ \Bar{A}_h) V\gth{g}{}^\dag\Big] \bigg), 
\end{align}{}
\end{widetext}
where we write $[X,Y]$ for the commutator of $X$ and $Y$.

Similarly to the approach in Section~\ref{ap:subsec:Control matrix}, we transform the control propagator into the eigenspace of $H_c\gth{g}{}$ and make use of its diagonal form given in Eq.~\eqref{ap:eq:U_bar}. Under consideration of the cyclic properties of a trace, and writing $\Delta\omega_{nm} = \omega_n\gth{g}{} - \omega_m\gth{g}{}$, the calculation can be further carried out as
\begin{widetext}
\begin{align}
    \frac{\partial \ctrlmat_{\alpha j}\gth{g}{}(\Delta t)}{\partial u_h (t_{g^\prime})} 
    &= \im \delta_{gg^\prime} s_\alpha\gth{g}{} \sum_{mn} \e^{\im \Delta\omega_{nm}\Delta t} \cdot \Bar{B}_{\alpha,nm} [\Bar{C_j}, \mathds{M}\gth{g}{}(t) \circ \Bar{A}_h]_{mn}.
\end{align}
\end{widetext}
Transforming the latter into frequency domain requires an integration over time. Since $s_\alpha\gth{g}{}$ and $\mc{B}_\alpha$ are time-independent within the regarded time step, this integration lies in
\begin{widetext}
\begin{subequations}
\begin{align}
    \mathds{K}_{pq}\gth{mn}{} &= \int_0^{\Delta t_g} dt \exp\left( \im \left( \omega + \Delta\omega_{nm} \right) t \right) \left[\mathds{M}\gth{g}{}(t)\right]_{pq} \nonumber\\
    &=\delta_{pq} \left( \frac{\Delta t_g\cdot\exp(\im(\omega + \Delta\omega_{nm})\Delta t_g)}{\im(\omega + \Delta\omega_{nm})} + \frac{\exp(\im(\omega + \Delta\omega_{nm})\Delta t_g) - 1}{(\omega + \Delta\omega_{nm})^2}\right)\nonumber\\
    &\hspace{0.4cm}+  \frac{1-\delta_{pq}}{\im\Delta\omega_{pq}} \cdot \left(\frac{\exp(\im(\omega + \Delta\omega_{nm} + \Delta\omega_{pq})\Delta t_g) - 1}{\im(\omega + \Delta\omega_{nm} + \Delta\omega_{pq})} - \frac{\exp(\im(\omega + \Delta\omega_{nm})\Delta t_g) - 1}{\im(\omega + \Delta\omega_{nm})}\right)
\end{align}
\text{if $\omega_p\gth{g}{} \neq \omega_q\gth{g}{}$ and}
\begin{align}
    \mathds{K}_{pq}\gth{mn}{} &= \frac{\Delta t_g\cdot\exp(\im(\omega + \Delta\omega_{nm})\Delta t_g)}{\im(\omega + \Delta\omega_{nm})} + \frac{\exp(\im(\omega + \Delta\omega_{nm})\Delta t_g) - 1}{(\omega + \Delta\omega_{nm})^2}
\end{align}
\text{otherwise.}
\end{subequations}
\end{widetext}
Using this result, we can first evaluate the transformed commutator element-wise to
\begin{align}
    \mathds{N}_{mn} &= \left[ \Bar{C}_j, \mathds{K}\gth{mn}{}\circ \Bar{A}_h \right]_{mn}, \label{ap:eq:N_mat}
\end{align}
and consequently derive a concise formula for the gradient of the control matrix at time step $g$ in the frequency domain as 
\begin{align}
    \pdv{\ctrlmat_{\alpha j}\gth{g}{}(\omega)}{u_h(t_{g^\prime})} &= \im\delta_{gg^\prime} s_\alpha\gth{g}{} \tr \left( \Bar{B}_{\alpha}\cdot \mathds{N} \right). \label{ap:eq:dR_g}
\end{align}

Inserting Eqs.~\eqref{ap:eq:R_g}, \eqref{ap:eq:dL} and \eqref{ap:eq:dR_g} into Eq.~\eqref{ap:eq:dR} gives the analytic derivative of the total control matrix. The latter result can then be used within Eq.~\eqref{eq:dF} to calculate analytic filter function derivatives for a general pulse sequence.

%--------------------------------------------------------
\section{\label{ap:AdditionalGraphics}Supplement information on software implementation and computational complexity}
%--------------------------------------------------------

%---------------------------------------
\subsection{Structure of implementation}
\begin{figure*}%[H]
    \centering
    \includegraphics[width=\textwidth]{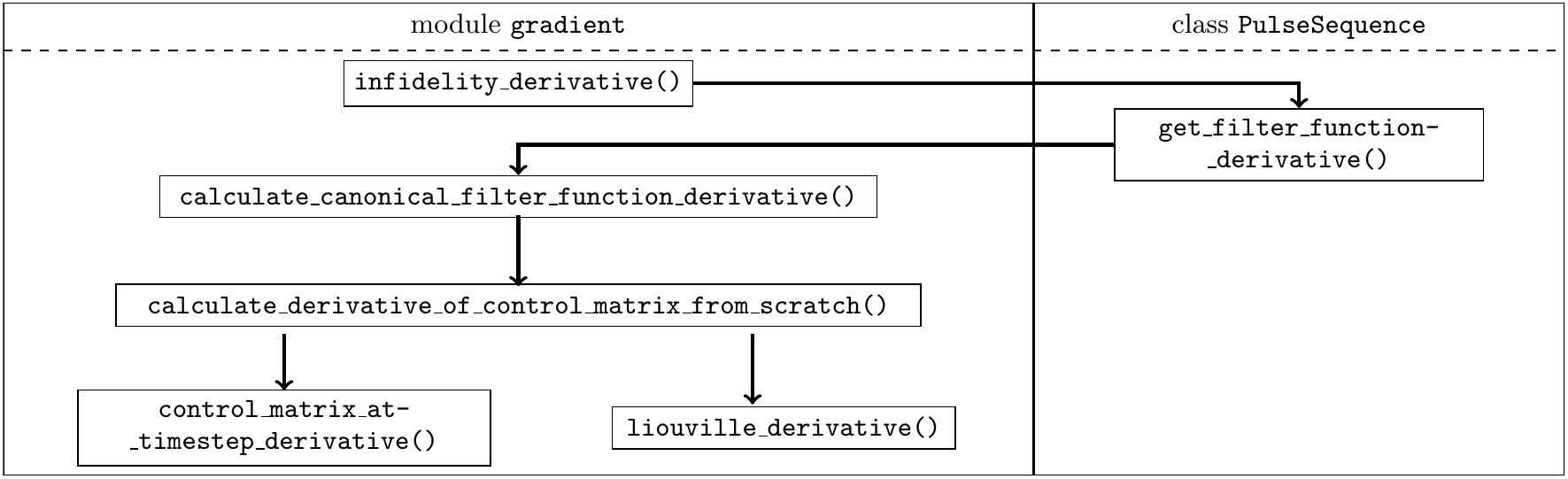}
    \caption{\label{ap:fig:flowchart}Structure of the implemented module \texttt{gradient} and newly added method in the \texttt{PulseSequence} class. The chart shows, which functions are called within each function. An arrow pointing from function \textit{A} to function \textit{B} indicates that \textit{A} calls \textit{B}.}
\end{figure*}

%---------------------------------------
In order to extend the \filterfunctions software package by filter function derivatives, the module \texttt{gradient} was implemented. Additionally, a method for directly obtaining filter function derivatives from a given pulse sequence was added to the \texttt{PulseSequence} class. The overall structure is illustrated in Fig.~\ref{ap:fig:flowchart}.

%--------------------------------------------------------
\subsection{Verification of linear run time dependency}
\begin{figure}%[H]
    \centering
    \includegraphics[width=\the\columnwidth]{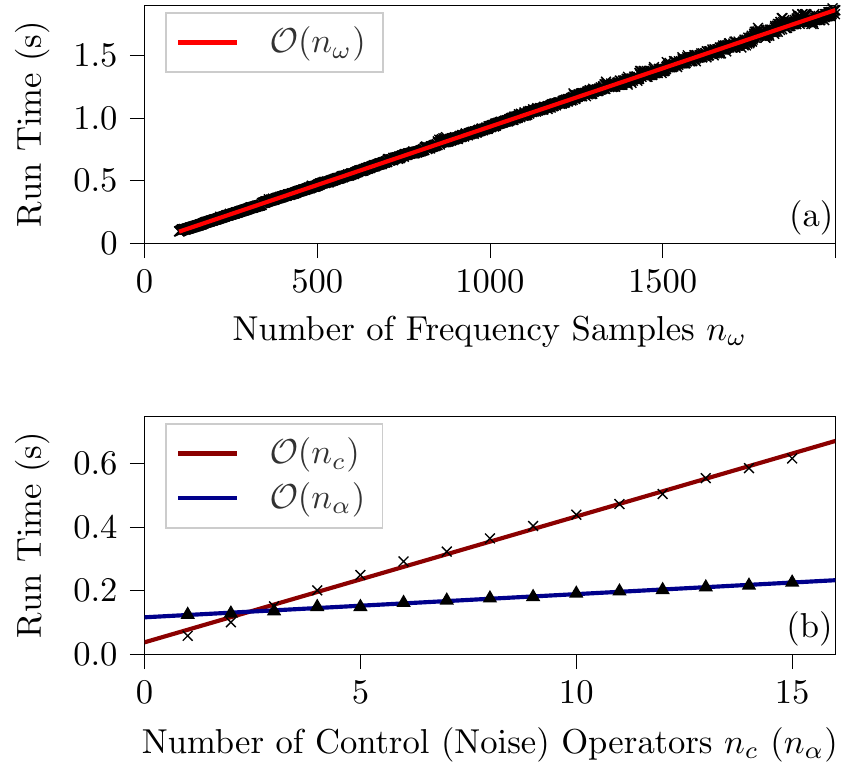}
    \caption{\label{ap:fig:runtimelinear}Linear run time behavior in $n_\omega$, $n_c$ and $n_\alpha$ of computing the infidelity gradient for random pulse sequences of dimension $d=4$ and $n_{\Delta t}=3$ time steps. (a) shows the run time plot for various $n_\omega$. To this end, a random pulse with $n_c=n_\alpha=2$ control and noise contributions, and $n_d=1$ drift contribution was generated. (b) shows the run time plot for various $n_c$ ($n_\alpha$). The plot shows the median data of 100 randomly generated pulse sequences with $n_\omega=200$, and $n_\alpha\;(n_c)=2$ depending on which parameter was fixed.}
\end{figure}

%--------------------------------------------------------
Within the run time analysis, the theoretically predicted linear dependency on the number of frequency samples, control operators, and noise operators could clearly be verified by means of asymptotic plots. Figure~\ref{ap:fig:runtimelinear} displays the analysis' results.

%----------------------------------------------------------------
\section{\label{sec:ComparisonNM}Supplement information on the comparison to Nelder-Mead method}
%----------------------------------------------------------------

In the following we will elaborate more on the details of the optimizations used to compare the use analytical gradients to a gradient free method. For better comparability, both optimization algorithms were started from the same initial pulses.

\begin{figure}
     \centering
     \includegraphics[width=\the\columnwidth]{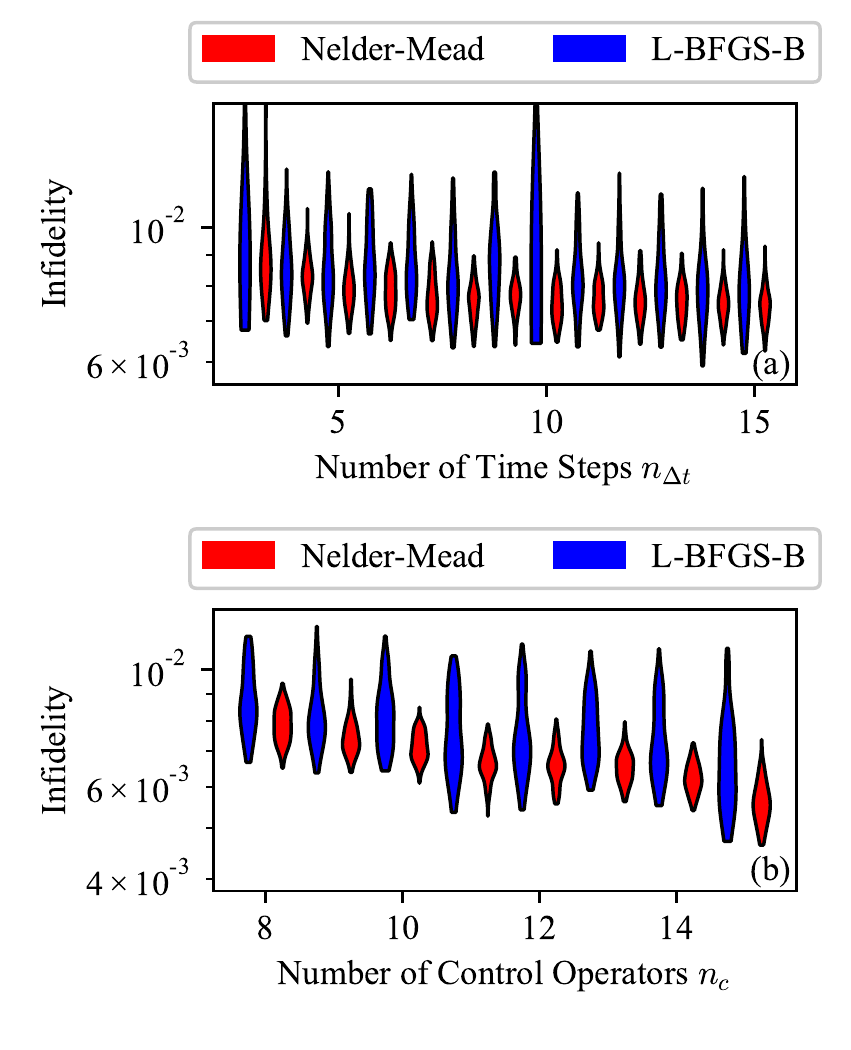}
     \caption{\label{fig:final_infid}Violin plot of the distribution of final total infidelities $\mc{I}$ of the optimizations (see Fig.~\ref{fig:convergence_time}). Each pair of a blue (L-BFGS) and a red (Nelder-Mead) violin shows the distributions of final infidelities for one set of 100 optimization runs, belonging to the number around which they are slightly shifted. (a) The distribution of final infidelities is plotted as function of the number of control operators $n_{\Delta t}$, while the number of time steps is kept constant at $n_c = 8$. (b) Here as function of the number of control operators $n_c$, the number of time steps is constantly $n_{\Delta t} = 6$.}
\end{figure}

In Fig.~\ref{fig:final_infid} we plotted the distribution of the optimized pulses' infidelities found in the optimization corresponding to Fig.~\ref{fig:convergence_time}. We can see that both algorithms find similar minimal infidelities over 100 runs. From the fact that the distribution of final cost values is wider for the L-BFGS-B algorithm, we can deduct that the gradient based method is more prone to get stuck in local minima in the optimization space, while the Nelder-Mead algorithm appears to provide better global convergence.

\begin{figure}
    \centering
    \includegraphics[width=\the\columnwidth]{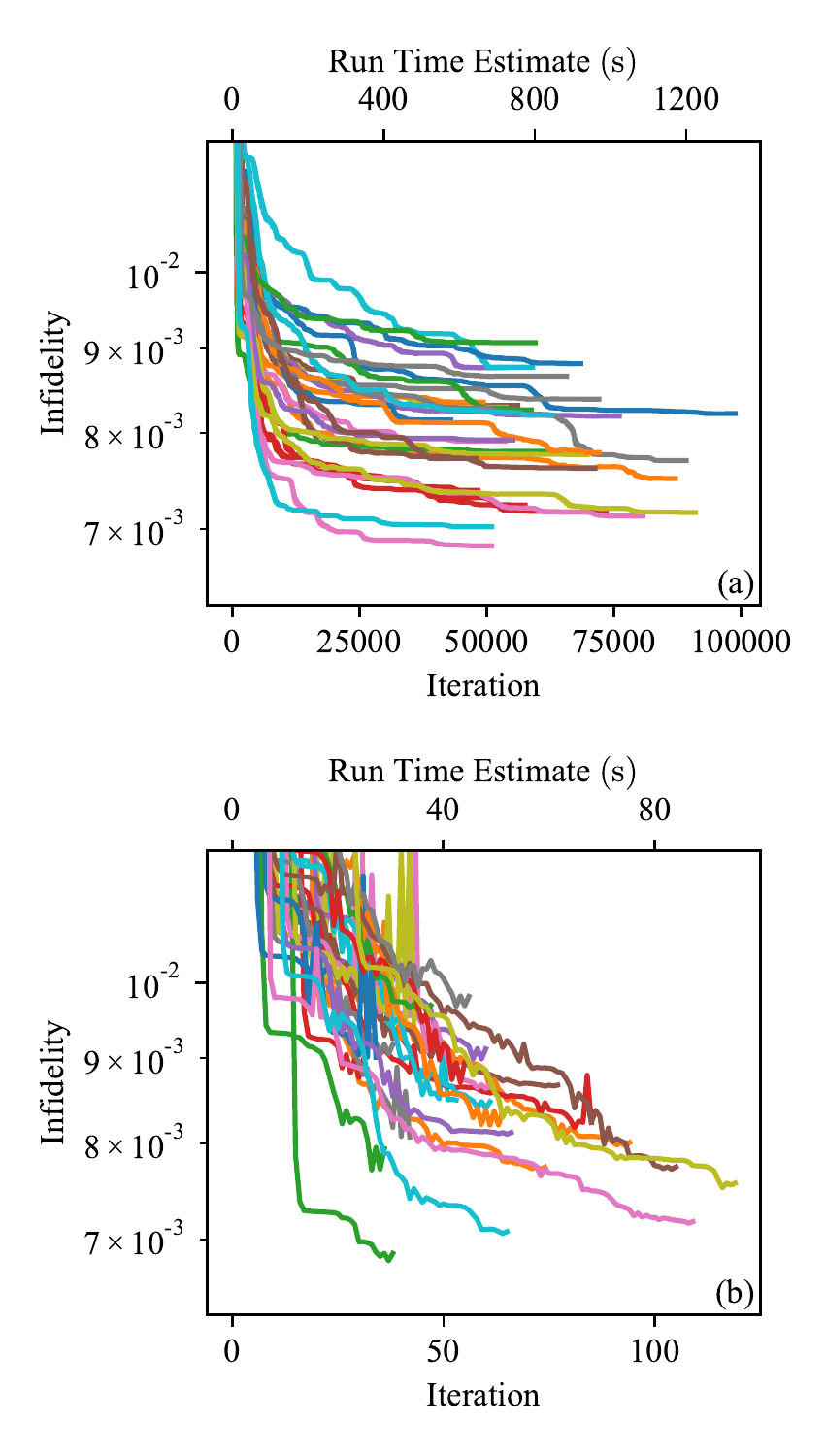}
    \caption{\label{fig:convergence_runs}Convergence behavior of the Nelder-Mead (a) and L-BFGS-B algorithm (b). Each plot shows the infidelity during 30 optimization runs as function the optimization algorithm's iteration. The runs were taken from the optimization with $n_{\Delta t} = 6$ and $n_c = 8$. We calculated the run time estimates by multiplying the iteration scale with the average duration of an iteration.}
\end{figure}

In the convergence plots in Fig.~\ref{fig:convergence_runs}, it can be seen that the gradient-based method requires far less iterations than the Nealder-Mead algorithm. The plateaus in the plots, where the algorithm reduces the Infidelity only marginally over several iterations, can be interpreted as features in the optimization landscape with the approximate form of local minima. We also observe, that the termination condition favors neither algorithm as both do not spend an excessive amount of iterations for minor improvements.

\begin{figure}[H]
    \centering
    \includegraphics[width=\the\columnwidth]{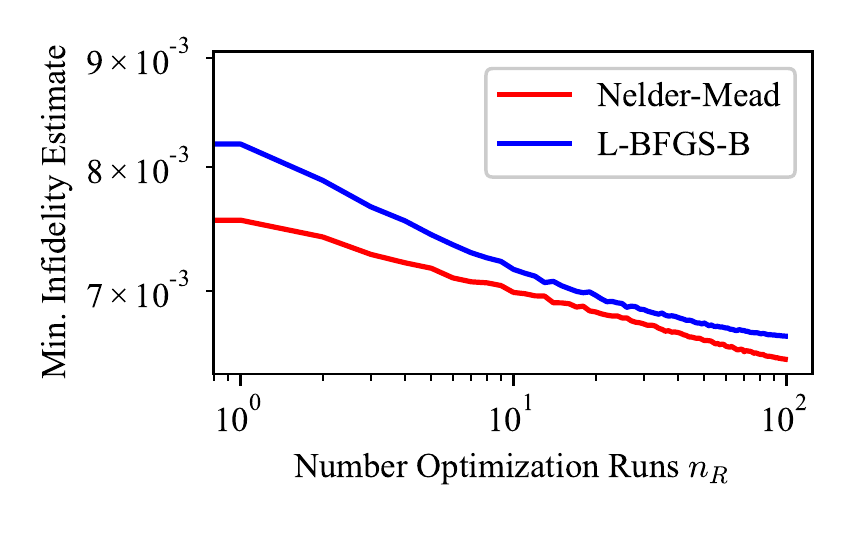}
    \caption{\label{fig:meta_convergence_runs} Estimation of the minimal infidelity achieved in $n_R$ optimization runs with different initial conditions. The estimate was calculated by averaging over the minimum of $n_R$ randomly drawn samples from the distribution plotted in Fig.~\ref{fig:final_infid} at $n_c = 8$.}
\end{figure}

To compare the convergence of the algorithms with respect to the initial parameters, we plotted in Fig.~\ref{fig:meta_convergence_runs} the expectation value of the infidelity as function of the number of optimization runs $n_R$. This quantifies the infidelity of the best pulse found by starting the optimization $n_R$ times with different initial pulse values. To find a pulse of similar infidelity on average, the L-BFGS-B algorithm needs to be restarted roughly twice as many times as the Nelder-Mead algorithm.

For the numerical pulse optimization, we used the open source python package \texttt{qopt} \cite{qopt2020, Teske2020}. 

%------------------------
%\bibliography{main.bib}

\begin{thebibliography}{46}%
\makeatletter
\providecommand \@ifxundefined [1]{%
 \@ifx{#1\undefined}
}%
\providecommand \@ifnum [1]{%
 \ifnum #1\expandafter \@firstoftwo
 \else \expandafter \@secondoftwo
 \fi
}%
\providecommand \@ifx [1]{%
 \ifx #1\expandafter \@firstoftwo
 \else \expandafter \@secondoftwo
 \fi
}%
\providecommand \natexlab [1]{#1}%
\providecommand \enquote  [1]{``#1''}%
\providecommand \bibnamefont  [1]{#1}%
\providecommand \bibfnamefont [1]{#1}%
\providecommand \citenamefont [1]{#1}%
\providecommand \href@noop [0]{\@secondoftwo}%
\providecommand \href [0]{\begingroup \@sanitize@url \@href}%
\providecommand \@href[1]{\@@startlink{#1}\@@href}%
\providecommand \@@href[1]{\endgroup#1\@@endlink}%
\providecommand \@sanitize@url [0]{\catcode `\\12\catcode `\$12\catcode
  `\&12\catcode `\#12\catcode `\^12\catcode `\_12\catcode `\%12\relax}%
\providecommand \@@startlink[1]{}%
\providecommand \@@endlink[0]{}%
\providecommand \url  [0]{\begingroup\@sanitize@url \@url }%
\providecommand \@url [1]{\endgroup\@href {#1}{\urlprefix }}%
\providecommand \urlprefix  [0]{URL }%
\providecommand \Eprint [0]{\href }%
\providecommand \doibase [0]{https://doi.org/}%
\providecommand \selectlanguage [0]{\@gobble}%
\providecommand \bibinfo  [0]{\@secondoftwo}%
\providecommand \bibfield  [0]{\@secondoftwo}%
\providecommand \translation [1]{[#1]}%
\providecommand \BibitemOpen [0]{}%
\providecommand \bibitemStop [0]{}%
\providecommand \bibitemNoStop [0]{.\EOS\space}%
\providecommand \EOS [0]{\spacefactor3000\relax}%
\providecommand \BibitemShut  [1]{\csname bibitem#1\endcsname}%
\let\auto@bib@innerbib\@empty
%</preamble>
\bibitem [{\citenamefont {Nielsen}\ and\ \citenamefont
  {Chuang}(2010)}]{Nielsen2010c}%
  \BibitemOpen
  \bibfield  {author} {\bibinfo {author} {\bibfnamefont {M.~A.}\ \bibnamefont
  {Nielsen}}\ and\ \bibinfo {author} {\bibfnamefont {I.~L.}\ \bibnamefont
  {Chuang}},\ }\href@noop {} {\emph {\bibinfo {title} {Quantum Computation and
  Quantum Information}}}\ (\bibinfo  {publisher} {Cambridge University Press},\
  \bibinfo {year} {2010})\BibitemShut {NoStop}%
\bibitem [{\citenamefont {Harrow}\ and\ \citenamefont
  {Montanaro}(2017)}]{Harrow2017}%
  \BibitemOpen
  \bibfield  {author} {\bibinfo {author} {\bibfnamefont {A.~W.}\ \bibnamefont
  {Harrow}}\ and\ \bibinfo {author} {\bibfnamefont {A.}~\bibnamefont
  {Montanaro}},\ }\bibfield  {title} {\bibinfo {title} {{Quantum computational
  supremacy}},\ }\href {https://doi.org/10.1038/nature23458} {\bibfield
  {journal} {\bibinfo  {journal} {Nature}\ }\textbf {\bibinfo {volume} {549}},\
  \bibinfo {pages} {203} (\bibinfo {year} {2017})}\BibitemShut {NoStop}%
\bibitem [{\citenamefont {Unruh}(1995)}]{Unruh1995}%
  \BibitemOpen
  \bibfield  {author} {\bibinfo {author} {\bibfnamefont {W.~G.}\ \bibnamefont
  {Unruh}},\ }\bibfield  {title} {\bibinfo {title} {Maintaining coherence in
  quantum computers},\ }\href {https://doi.org/10.1103/PhysRevA.51.992}
  {\bibfield  {journal} {\bibinfo  {journal} {Phys. Rev. A}\ }\textbf {\bibinfo
  {volume} {51}},\ \bibinfo {pages} {992} (\bibinfo {year} {1995})}\BibitemShut
  {NoStop}%
\bibitem [{\citenamefont {Preskill}(2018)}]{Preskill2018}%
  \BibitemOpen
  \bibfield  {author} {\bibinfo {author} {\bibfnamefont {J.}~\bibnamefont
  {Preskill}},\ }\bibfield  {title} {\bibinfo {title} {Quantum {C}omputing in
  the {NISQ} era and beyond},\ }\href
  {https://doi.org/10.22331/q-2018-08-06-79} {\bibfield  {journal} {\bibinfo
  {journal} {{Quantum}}\ }\textbf {\bibinfo {volume} {2}},\ \bibinfo {pages}
  {79} (\bibinfo {year} {2018})}\BibitemShut {NoStop}%
\bibitem [{\citenamefont {Glaser}\ \emph {et~al.}(2015)\citenamefont {Glaser},
  \citenamefont {Boscain}, \citenamefont {Calarco}, \citenamefont {Koch},
  \citenamefont {K{\"{o}}ckenberger}, \citenamefont {Kosloff}, \citenamefont
  {Kuprov}, \citenamefont {Luy}, \citenamefont {Schirmer}, \citenamefont
  {Schulte-Herbr{\"{u}}ggen}, \citenamefont {Sugny},\ and\ \citenamefont
  {Wilhelm}}]{Glaser2015}%
  \BibitemOpen
  \bibfield  {author} {\bibinfo {author} {\bibfnamefont {S.~J.}\ \bibnamefont
  {Glaser}}, \bibinfo {author} {\bibfnamefont {U.}~\bibnamefont {Boscain}},
  \bibinfo {author} {\bibfnamefont {T.}~\bibnamefont {Calarco}}, \bibinfo
  {author} {\bibfnamefont {C.~P.}\ \bibnamefont {Koch}}, \bibinfo {author}
  {\bibfnamefont {W.}~\bibnamefont {K{\"{o}}ckenberger}}, \bibinfo {author}
  {\bibfnamefont {R.}~\bibnamefont {Kosloff}}, \bibinfo {author} {\bibfnamefont
  {I.}~\bibnamefont {Kuprov}}, \bibinfo {author} {\bibfnamefont
  {B.}~\bibnamefont {Luy}}, \bibinfo {author} {\bibfnamefont {S.}~\bibnamefont
  {Schirmer}}, \bibinfo {author} {\bibfnamefont {T.}~\bibnamefont
  {Schulte-Herbr{\"{u}}ggen}}, \bibinfo {author} {\bibfnamefont
  {D.}~\bibnamefont {Sugny}},\ and\ \bibinfo {author} {\bibfnamefont {F.~K.}\
  \bibnamefont {Wilhelm}},\ }\bibfield  {title} {\bibinfo {title} {{Training
  Schr{\"{o}}dinger's cat: Quantum optimal control: Strategic report on current
  status, visions and goals for research in Europe}},\ }\href
  {https://doi.org/10.1140/epjd/e2015-60464-1} {\bibfield  {journal} {\bibinfo
  {journal} {Eur. Phys. J. D}\ }\textbf {\bibinfo {volume} {69}},\ \bibinfo
  {pages} {279} (\bibinfo {year} {2015})}\BibitemShut {NoStop}%
\bibitem [{\citenamefont {Peirce}\ \emph {et~al.}(1988)\citenamefont {Peirce},
  \citenamefont {Dahleh},\ and\ \citenamefont {Rabitz}}]{Peirce1988}%
  \BibitemOpen
  \bibfield  {author} {\bibinfo {author} {\bibfnamefont {A.~P.}\ \bibnamefont
  {Peirce}}, \bibinfo {author} {\bibfnamefont {M.~A.}\ \bibnamefont {Dahleh}},\
  and\ \bibinfo {author} {\bibfnamefont {H.}~\bibnamefont {Rabitz}},\
  }\bibfield  {title} {\bibinfo {title} {Optimal control of quantum-mechanical
  systems: Existence, numerical approximation, and applications},\ }\href
  {https://doi.org/10.1103/PhysRevA.37.4950} {\bibfield  {journal} {\bibinfo
  {journal} {Phys. Rev. A}\ }\textbf {\bibinfo {volume} {37}},\ \bibinfo
  {pages} {4950} (\bibinfo {year} {1988})}\BibitemShut {NoStop}%
\bibitem [{\citenamefont {Rabitz}(2009)}]{Rabitz2009}%
  \BibitemOpen
  \bibfield  {author} {\bibinfo {author} {\bibfnamefont {H.}~\bibnamefont
  {Rabitz}},\ }\bibfield  {title} {\bibinfo {title} {{Focus on Quantum
  Control}},\ }\href {https://doi.org/10.1088/1367-2630/11/10/105030}
  {\bibfield  {journal} {\bibinfo  {journal} {New J. Phys}\ }\textbf {\bibinfo
  {volume} {11}},\ \bibinfo {pages} {105030} (\bibinfo {year}
  {2009})}\BibitemShut {NoStop}%
\bibitem [{\citenamefont {Wang}\ \emph {et~al.}(2012)\citenamefont {Wang},
  \citenamefont {Bishop}, \citenamefont {Kestner}, \citenamefont {Barnes},
  \citenamefont {Sun},\ and\ \citenamefont {Sarma}}]{Wang2012}%
  \BibitemOpen
  \bibfield  {author} {\bibinfo {author} {\bibfnamefont {X.}~\bibnamefont
  {Wang}}, \bibinfo {author} {\bibfnamefont {L.~S.}\ \bibnamefont {Bishop}},
  \bibinfo {author} {\bibfnamefont {J.~P.}\ \bibnamefont {Kestner}}, \bibinfo
  {author} {\bibfnamefont {E.}~\bibnamefont {Barnes}}, \bibinfo {author}
  {\bibfnamefont {K.}~\bibnamefont {Sun}},\ and\ \bibinfo {author}
  {\bibfnamefont {D.}~\bibnamefont {Sarma}},\ }\bibfield  {title} {\bibinfo
  {title} {{Composite pulses for robust universal control of singlet-triplet
  qubits}},\ }\href {https://doi.org/10.1038/ncomms2003} {\bibfield  {journal}
  {\bibinfo  {journal} {Nat. Commun.}\ }\textbf {\bibinfo {volume} {3}},\
  \bibinfo {pages} {997} (\bibinfo {year} {2012})}\BibitemShut {NoStop}%
\bibitem [{\citenamefont {Wang}\ \emph
  {et~al.}(2014{\natexlab{a}})\citenamefont {Wang}, \citenamefont {Bishop},
  \citenamefont {Barnes}, \citenamefont {Kestner},\ and\ \citenamefont
  {Sarma}}]{Wang2014}%
  \BibitemOpen
  \bibfield  {author} {\bibinfo {author} {\bibfnamefont {X.}~\bibnamefont
  {Wang}}, \bibinfo {author} {\bibfnamefont {L.~S.}\ \bibnamefont {Bishop}},
  \bibinfo {author} {\bibfnamefont {E.}~\bibnamefont {Barnes}}, \bibinfo
  {author} {\bibfnamefont {J.~P.}\ \bibnamefont {Kestner}},\ and\ \bibinfo
  {author} {\bibfnamefont {S.~D.}\ \bibnamefont {Sarma}},\ }\bibfield  {title}
  {\bibinfo {title} {Robust quantum gates for singlet-triplet spin qubits using
  composite pulses},\ }\href {https://doi.org/10.1103/PhysRevA.89.022310}
  {\bibfield  {journal} {\bibinfo  {journal} {Phys. Rev. A}\ }\textbf {\bibinfo
  {volume} {89}},\ \bibinfo {pages} {022310} (\bibinfo {year}
  {2014}{\natexlab{a}})}\BibitemShut {NoStop}%
\bibitem [{\citenamefont {Wang}\ \emph
  {et~al.}(2014{\natexlab{b}})\citenamefont {Wang}, \citenamefont
  {Calderon-Vargas}, \citenamefont {Rana}, \citenamefont {Kestner},
  \citenamefont {Barnes},\ and\ \citenamefont {Das~Sarma}}]{Wang2014a}%
  \BibitemOpen
  \bibfield  {author} {\bibinfo {author} {\bibfnamefont {X.}~\bibnamefont
  {Wang}}, \bibinfo {author} {\bibfnamefont {F.~A.}\ \bibnamefont
  {Calderon-Vargas}}, \bibinfo {author} {\bibfnamefont {M.~S.}\ \bibnamefont
  {Rana}}, \bibinfo {author} {\bibfnamefont {J.~P.}\ \bibnamefont {Kestner}},
  \bibinfo {author} {\bibfnamefont {E.}~\bibnamefont {Barnes}},\ and\ \bibinfo
  {author} {\bibfnamefont {S.}~\bibnamefont {Das~Sarma}},\ }\bibfield  {title}
  {\bibinfo {title} {Noise-compensating pulses for electrostatically controlled
  silicon spin qubits},\ }\href {https://doi.org/10.1103/PhysRevB.90.155306}
  {\bibfield  {journal} {\bibinfo  {journal} {Phys. Rev. B}\ }\textbf {\bibinfo
  {volume} {90}},\ \bibinfo {pages} {155306} (\bibinfo {year}
  {2014}{\natexlab{b}})}\BibitemShut {NoStop}%
\bibitem [{\citenamefont {Yang}\ and\ \citenamefont {Wang}(2016)}]{Yang2016}%
  \BibitemOpen
  \bibfield  {author} {\bibinfo {author} {\bibfnamefont {X.~C.}\ \bibnamefont
  {Yang}}\ and\ \bibinfo {author} {\bibfnamefont {X.}~\bibnamefont {Wang}},\
  }\bibfield  {title} {\bibinfo {title} {{Noise filtering of composite pulses
  for singlet-triplet qubits}},\ }\href {https://doi.org/10.1038/srep28996}
  {\bibfield  {journal} {\bibinfo  {journal} {Sci. Rep.}\ }\textbf {\bibinfo
  {volume} {6}},\ \bibinfo {pages} {28996} (\bibinfo {year}
  {2016})}\BibitemShut {NoStop}%
\bibitem [{\citenamefont {Wellstood}\ \emph {et~al.}(1987)\citenamefont
  {Wellstood}, \citenamefont {Urbina},\ and\ \citenamefont
  {Clarke}}]{Wellstood1987}%
  \BibitemOpen
  \bibfield  {author} {\bibinfo {author} {\bibfnamefont {F.~C.}\ \bibnamefont
  {Wellstood}}, \bibinfo {author} {\bibfnamefont {C.}~\bibnamefont {Urbina}},\
  and\ \bibinfo {author} {\bibfnamefont {J.}~\bibnamefont {Clarke}},\
  }\bibfield  {title} {\bibinfo {title} {Low‐frequency noise in dc
  superconducting quantum interference devices below 1 k},\ }\href
  {https://doi.org/10.1063/1.98041} {\bibfield  {journal} {\bibinfo  {journal}
  {Appl. Phys. Lett.}\ }\textbf {\bibinfo {volume} {50}},\ \bibinfo {pages}
  {772} (\bibinfo {year} {1987})}\BibitemShut {NoStop}%
\bibitem [{\citenamefont {Bylander}\ \emph {et~al.}(2011)\citenamefont
  {Bylander}, \citenamefont {Gustavsson}, \citenamefont {Yan}, \citenamefont
  {Yoshihara}, \citenamefont {Harrabi}, \citenamefont {Fitch}, \citenamefont
  {Cory}, \citenamefont {Nakamura}, \citenamefont {Tsai},\ and\ \citenamefont
  {Oliver}}]{Bylander2011}%
  \BibitemOpen
  \bibfield  {author} {\bibinfo {author} {\bibfnamefont {J.}~\bibnamefont
  {Bylander}}, \bibinfo {author} {\bibfnamefont {S.}~\bibnamefont
  {Gustavsson}}, \bibinfo {author} {\bibfnamefont {F.}~\bibnamefont {Yan}},
  \bibinfo {author} {\bibfnamefont {F.}~\bibnamefont {Yoshihara}}, \bibinfo
  {author} {\bibfnamefont {K.}~\bibnamefont {Harrabi}}, \bibinfo {author}
  {\bibfnamefont {G.}~\bibnamefont {Fitch}}, \bibinfo {author} {\bibfnamefont
  {D.~G.}\ \bibnamefont {Cory}}, \bibinfo {author} {\bibfnamefont
  {Y.}~\bibnamefont {Nakamura}}, \bibinfo {author} {\bibfnamefont {J.~S.}\
  \bibnamefont {Tsai}},\ and\ \bibinfo {author} {\bibfnamefont {W.~D.}\
  \bibnamefont {Oliver}},\ }\bibfield  {title} {\bibinfo {title} {{Noise
  spectroscopy through dynamical decoupling with a superconducting flux
  qubit}},\ }\href {https://doi.org/10.1038/nphys1994} {\bibfield  {journal}
  {\bibinfo  {journal} {Nat. Phys.}\ }\textbf {\bibinfo {volume} {7}},\
  \bibinfo {pages} {565} (\bibinfo {year} {2011})}\BibitemShut {NoStop}%
\bibitem [{\citenamefont {{Drung}}\ \emph {et~al.}(2011)\citenamefont
  {{Drung}}, \citenamefont {{Beyer}}, \citenamefont {{Storm}}, \citenamefont
  {{Peters}},\ and\ \citenamefont {{Schurig}}}]{Drung2011}%
  \BibitemOpen
  \bibfield  {author} {\bibinfo {author} {\bibfnamefont {D.}~\bibnamefont
  {{Drung}}}, \bibinfo {author} {\bibfnamefont {J.}~\bibnamefont {{Beyer}}},
  \bibinfo {author} {\bibfnamefont {J.}~\bibnamefont {{Storm}}}, \bibinfo
  {author} {\bibfnamefont {M.}~\bibnamefont {{Peters}}},\ and\ \bibinfo
  {author} {\bibfnamefont {T.}~\bibnamefont {{Schurig}}},\ }\bibfield  {title}
  {\bibinfo {title} {Investigation of low-frequency excess flux noise in dc
  squids at mk temperatures},\ }\href
  {https://doi.org/10.1109/TASC.2010.2084054} {\bibfield  {journal} {\bibinfo
  {journal} {IEEE Trans. Appl. Supercond.}\ }\textbf {\bibinfo {volume} {21}},\
  \bibinfo {pages} {340} (\bibinfo {year} {2011})}\BibitemShut {NoStop}%
\bibitem [{\citenamefont {Anton}\ \emph {et~al.}(2012)\citenamefont {Anton},
  \citenamefont {M\"uller}, \citenamefont {Birenbaum}, \citenamefont
  {O'Kelley}, \citenamefont {Fefferman}, \citenamefont {Golubev}, \citenamefont
  {Hilton}, \citenamefont {Cho}, \citenamefont {Irwin}, \citenamefont
  {Wellstood}, \citenamefont {Sch\"on}, \citenamefont {Shnirman},\ and\
  \citenamefont {Clarke}}]{Anton2012}%
  \BibitemOpen
  \bibfield  {author} {\bibinfo {author} {\bibfnamefont {S.~M.}\ \bibnamefont
  {Anton}}, \bibinfo {author} {\bibfnamefont {C.}~\bibnamefont {M\"uller}},
  \bibinfo {author} {\bibfnamefont {J.~S.}\ \bibnamefont {Birenbaum}}, \bibinfo
  {author} {\bibfnamefont {S.~R.}\ \bibnamefont {O'Kelley}}, \bibinfo {author}
  {\bibfnamefont {A.~D.}\ \bibnamefont {Fefferman}}, \bibinfo {author}
  {\bibfnamefont {D.~S.}\ \bibnamefont {Golubev}}, \bibinfo {author}
  {\bibfnamefont {G.~C.}\ \bibnamefont {Hilton}}, \bibinfo {author}
  {\bibfnamefont {H.-M.}\ \bibnamefont {Cho}}, \bibinfo {author} {\bibfnamefont
  {K.~D.}\ \bibnamefont {Irwin}}, \bibinfo {author} {\bibfnamefont {F.~C.}\
  \bibnamefont {Wellstood}}, \bibinfo {author} {\bibfnamefont {G.}~\bibnamefont
  {Sch\"on}}, \bibinfo {author} {\bibfnamefont {A.}~\bibnamefont {Shnirman}},\
  and\ \bibinfo {author} {\bibfnamefont {J.}~\bibnamefont {Clarke}},\
  }\bibfield  {title} {\bibinfo {title} {Pure dephasing in flux qubits due to
  flux noise with spectral density scaling as $1/{f}^{\ensuremath{\alpha}}$},\
  }\href {https://doi.org/10.1103/PhysRevB.85.224505} {\bibfield  {journal}
  {\bibinfo  {journal} {Phys. Rev. B}\ }\textbf {\bibinfo {volume} {85}},\
  \bibinfo {pages} {224505} (\bibinfo {year} {2012})}\BibitemShut {NoStop}%
\bibitem [{\citenamefont {Kuhlmann}\ \emph {et~al.}(2013)\citenamefont
  {Kuhlmann}, \citenamefont {Houel}, \citenamefont {Ludwig}, \citenamefont
  {Greuter}, \citenamefont {Reuter}, \citenamefont {Wieck}, \citenamefont
  {Poggio},\ and\ \citenamefont {Warburton}}]{Kuhlmann2013}%
  \BibitemOpen
  \bibfield  {author} {\bibinfo {author} {\bibfnamefont {A.~V.}\ \bibnamefont
  {Kuhlmann}}, \bibinfo {author} {\bibfnamefont {J.}~\bibnamefont {Houel}},
  \bibinfo {author} {\bibfnamefont {A.}~\bibnamefont {Ludwig}}, \bibinfo
  {author} {\bibfnamefont {L.}~\bibnamefont {Greuter}}, \bibinfo {author}
  {\bibfnamefont {D.}~\bibnamefont {Reuter}}, \bibinfo {author} {\bibfnamefont
  {A.~D.}\ \bibnamefont {Wieck}}, \bibinfo {author} {\bibfnamefont
  {M.}~\bibnamefont {Poggio}},\ and\ \bibinfo {author} {\bibfnamefont {R.~J.}\
  \bibnamefont {Warburton}},\ }\bibfield  {title} {\bibinfo {title} {{Charge
  noise and spin noise in a semiconductor quantum device}},\ }\href
  {https://doi.org/10.1038/nphys2688} {\bibfield  {journal} {\bibinfo
  {journal} {Nat. Phys.}\ }\textbf {\bibinfo {volume} {9}},\ \bibinfo {pages}
  {570} (\bibinfo {year} {2013})}\BibitemShut {NoStop}%
\bibitem [{\citenamefont {Yoneda}\ \emph {et~al.}(2018)\citenamefont {Yoneda},
  \citenamefont {Takeda}, \citenamefont {Otsuka}, \citenamefont {Nakajima},
  \citenamefont {Delbecq}, \citenamefont {Allison}, \citenamefont {Honda},
  \citenamefont {Kodera}, \citenamefont {Oda}, \citenamefont {Hoshi},
  \citenamefont {Usami}, \citenamefont {Itoh},\ and\ \citenamefont
  {Tarucha}}]{Yoneda2018}%
  \BibitemOpen
  \bibfield  {author} {\bibinfo {author} {\bibfnamefont {J.}~\bibnamefont
  {Yoneda}}, \bibinfo {author} {\bibfnamefont {K.}~\bibnamefont {Takeda}},
  \bibinfo {author} {\bibfnamefont {T.}~\bibnamefont {Otsuka}}, \bibinfo
  {author} {\bibfnamefont {T.}~\bibnamefont {Nakajima}}, \bibinfo {author}
  {\bibfnamefont {M.~R.}\ \bibnamefont {Delbecq}}, \bibinfo {author}
  {\bibfnamefont {G.}~\bibnamefont {Allison}}, \bibinfo {author} {\bibfnamefont
  {T.}~\bibnamefont {Honda}}, \bibinfo {author} {\bibfnamefont
  {T.}~\bibnamefont {Kodera}}, \bibinfo {author} {\bibfnamefont
  {S.}~\bibnamefont {Oda}}, \bibinfo {author} {\bibfnamefont {Y.}~\bibnamefont
  {Hoshi}}, \bibinfo {author} {\bibfnamefont {N.}~\bibnamefont {Usami}},
  \bibinfo {author} {\bibfnamefont {K.~M.}\ \bibnamefont {Itoh}},\ and\
  \bibinfo {author} {\bibfnamefont {S.}~\bibnamefont {Tarucha}},\ }\bibfield
  {title} {\bibinfo {title} {{A quantum-dot spin qubit with coherence limited
  by charge noise and fidelity higher than 99.9{\%}}},\ }\href
  {https://doi.org/10.1038/s41565-017-0014-x} {\bibfield  {journal} {\bibinfo
  {journal} {Nat. Nanotechnol.}\ }\textbf {\bibinfo {volume} {13}},\ \bibinfo
  {pages} {102} (\bibinfo {year} {2018})}\BibitemShut {NoStop}%
\bibitem [{\citenamefont {Struck}\ \emph {et~al.}(2020)\citenamefont {Struck},
  \citenamefont {Hollmann}, \citenamefont {Schauer}, \citenamefont {Fedorets},
  \citenamefont {Schmidbauer}, \citenamefont {Sawano}, \citenamefont {Riemann},
  \citenamefont {Abrosimov}, \citenamefont {Cywi{\'{n}}ski}, \citenamefont
  {Bougeard},\ and\ \citenamefont {Schreiber}}]{Struck2020}%
  \BibitemOpen
  \bibfield  {author} {\bibinfo {author} {\bibfnamefont {T.}~\bibnamefont
  {Struck}}, \bibinfo {author} {\bibfnamefont {A.}~\bibnamefont {Hollmann}},
  \bibinfo {author} {\bibfnamefont {F.}~\bibnamefont {Schauer}}, \bibinfo
  {author} {\bibfnamefont {O.}~\bibnamefont {Fedorets}}, \bibinfo {author}
  {\bibfnamefont {A.}~\bibnamefont {Schmidbauer}}, \bibinfo {author}
  {\bibfnamefont {K.}~\bibnamefont {Sawano}}, \bibinfo {author} {\bibfnamefont
  {H.}~\bibnamefont {Riemann}}, \bibinfo {author} {\bibfnamefont {N.~V.}\
  \bibnamefont {Abrosimov}}, \bibinfo {author} {\bibfnamefont
  {{\L}.}~\bibnamefont {Cywi{\'{n}}ski}}, \bibinfo {author} {\bibfnamefont
  {D.}~\bibnamefont {Bougeard}},\ and\ \bibinfo {author} {\bibfnamefont
  {L.~R.}\ \bibnamefont {Schreiber}},\ }\bibfield  {title} {\bibinfo {title}
  {{Low-frequency spin qubit energy splitting noise in highly purified
  28Si/SiGe}},\ }\href {https://doi.org/10.1038/s41534-020-0276-2} {\bibfield
  {journal} {\bibinfo  {journal} {npj Quantum Inf.}\ }\textbf {\bibinfo
  {volume} {6}},\ \bibinfo {pages} {40} (\bibinfo {year} {2020})}\BibitemShut
  {NoStop}%
\bibitem [{\citenamefont {Green}\ \emph {et~al.}(2013)\citenamefont {Green},
  \citenamefont {Sastrawan}, \citenamefont {Uys},\ and\ \citenamefont
  {Biercuk}}]{Green2013}%
  \BibitemOpen
  \bibfield  {author} {\bibinfo {author} {\bibfnamefont {T.~J.}\ \bibnamefont
  {Green}}, \bibinfo {author} {\bibfnamefont {J.}~\bibnamefont {Sastrawan}},
  \bibinfo {author} {\bibfnamefont {H.}~\bibnamefont {Uys}},\ and\ \bibinfo
  {author} {\bibfnamefont {M.~J.}\ \bibnamefont {Biercuk}},\ }\bibfield
  {title} {\bibinfo {title} {Arbitrary quantum control of qubits in the
  presence of universal noise},\ }\href
  {https://doi.org/10.1088/1367-2630/15/9/095004} {\bibfield  {journal}
  {\bibinfo  {journal} {New J. Phys.}\ }\textbf {\bibinfo {volume} {15}},\
  \bibinfo {pages} {095004} (\bibinfo {year} {2013})}\BibitemShut {NoStop}%
\bibitem [{\citenamefont {Cerfontaine}\ \emph {et~al.}(2021)\citenamefont
  {Cerfontaine}, \citenamefont {Hangleiter},\ and\ \citenamefont
  {Bluhm}}]{Cerfontaine2020}%
  \BibitemOpen
  \bibfield  {author} {\bibinfo {author} {\bibfnamefont {P.}~\bibnamefont
  {Cerfontaine}}, \bibinfo {author} {\bibfnamefont {T.}~\bibnamefont
  {Hangleiter}},\ and\ \bibinfo {author} {\bibfnamefont {H.}~\bibnamefont
  {Bluhm}},\ }\href@noop {} {\bibinfo {title} {Filter functions for quantum
  processes under correlated noise}} (\bibinfo {year} {2021}),\ \Eprint
  {https://arxiv.org/abs/2103.02385} {arXiv:2103.02385 [quant-ph]} \BibitemShut
  {NoStop}%
\bibitem [{\citenamefont {Hangleiter}\ \emph
  {et~al.}(2021{\natexlab{a}})\citenamefont {Hangleiter}, \citenamefont
  {Cerfontaine},\ and\ \citenamefont {Bluhm}}]{Hangleiter2020}%
  \BibitemOpen
  \bibfield  {author} {\bibinfo {author} {\bibfnamefont {T.}~\bibnamefont
  {Hangleiter}}, \bibinfo {author} {\bibfnamefont {P.}~\bibnamefont
  {Cerfontaine}},\ and\ \bibinfo {author} {\bibfnamefont {H.}~\bibnamefont
  {Bluhm}},\ }\href@noop {} {\bibinfo {title} {Filter function formalism and
  software package to compute quantum processes of gate sequences for classical
  non-markovian noise}} (\bibinfo {year} {2021}{\natexlab{a}}),\ \Eprint
  {https://arxiv.org/abs/2103.02403} {arXiv:2103.02403 [quant-ph]} \BibitemShut
  {NoStop}%
\bibitem [{\citenamefont {Soare}\ \emph {et~al.}(2014)\citenamefont {Soare},
  \citenamefont {Ball}, \citenamefont {Hayes}, \citenamefont {Sastrawan},
  \citenamefont {Jarratt}, \citenamefont {Mcloughlin}, \citenamefont {Zhen},
  \citenamefont {Green},\ and\ \citenamefont {Biercuk}}]{Soare2014}%
  \BibitemOpen
  \bibfield  {author} {\bibinfo {author} {\bibfnamefont {A.}~\bibnamefont
  {Soare}}, \bibinfo {author} {\bibfnamefont {H.}~\bibnamefont {Ball}},
  \bibinfo {author} {\bibfnamefont {D.}~\bibnamefont {Hayes}}, \bibinfo
  {author} {\bibfnamefont {J.}~\bibnamefont {Sastrawan}}, \bibinfo {author}
  {\bibfnamefont {M.~C.}\ \bibnamefont {Jarratt}}, \bibinfo {author}
  {\bibfnamefont {J.~J.}\ \bibnamefont {Mcloughlin}}, \bibinfo {author}
  {\bibfnamefont {X.}~\bibnamefont {Zhen}}, \bibinfo {author} {\bibfnamefont
  {T.~J.}\ \bibnamefont {Green}},\ and\ \bibinfo {author} {\bibfnamefont
  {M.~J.}\ \bibnamefont {Biercuk}},\ }\bibfield  {title} {\bibinfo {title}
  {{Experimental noise filtering by quantum control}},\ }\href
  {https://doi.org/10.1038/nphys3115} {\bibfield  {journal} {\bibinfo
  {journal} {Nat. Phys.}\ }\textbf {\bibinfo {volume} {10}},\ \bibinfo {pages}
  {825} (\bibinfo {year} {2014})}\BibitemShut {NoStop}%
\bibitem [{\citenamefont {Nielsen}(2002)}]{Nielsen2002}%
  \BibitemOpen
  \bibfield  {author} {\bibinfo {author} {\bibfnamefont {M.~A.}\ \bibnamefont
  {Nielsen}},\ }\bibfield  {title} {\bibinfo {title} {A simple formula for the
  average gate fidelity of a quantum dynamical operation},\ }\href
  {https://doi.org/https://doi.org/10.1016/S0375-9601(02)01272-0} {\bibfield
  {journal} {\bibinfo  {journal} {Phys. Lett. A}\ }\textbf {\bibinfo {volume}
  {303}},\ \bibinfo {pages} {249} (\bibinfo {year} {2002})}\BibitemShut
  {NoStop}%
\bibitem [{\citenamefont {Green}\ \emph {et~al.}(2012)\citenamefont {Green},
  \citenamefont {Uys},\ and\ \citenamefont {Biercuk}}]{Green2012}%
  \BibitemOpen
  \bibfield  {author} {\bibinfo {author} {\bibfnamefont {T.}~\bibnamefont
  {Green}}, \bibinfo {author} {\bibfnamefont {H.}~\bibnamefont {Uys}},\ and\
  \bibinfo {author} {\bibfnamefont {M.~J.}\ \bibnamefont {Biercuk}},\
  }\bibfield  {title} {\bibinfo {title} {High-order noise filtering in
  nontrivial quantum logic gates},\ }\href
  {https://doi.org/10.1103/PhysRevLett.109.020501} {\bibfield  {journal}
  {\bibinfo  {journal} {Phys. Rev. Lett.}\ }\textbf {\bibinfo {volume} {109}},\
  \bibinfo {pages} {020501} (\bibinfo {year} {2012})}\BibitemShut {NoStop}%
\bibitem [{\citenamefont {Huang}\ and\ \citenamefont {Goan}(2017)}]{Huang2017}%
  \BibitemOpen
  \bibfield  {author} {\bibinfo {author} {\bibfnamefont {C.-H.}\ \bibnamefont
  {Huang}}\ and\ \bibinfo {author} {\bibfnamefont {H.-S.}\ \bibnamefont
  {Goan}},\ }\bibfield  {title} {\bibinfo {title} {Robust quantum gates for
  stochastic time-varying noise},\ }\href
  {https://doi.org/10.1103/PhysRevA.95.062325} {\bibfield  {journal} {\bibinfo
  {journal} {Phys. Rev. A}\ }\textbf {\bibinfo {volume} {95}},\ \bibinfo
  {pages} {062325} (\bibinfo {year} {2017})}\BibitemShut {NoStop}%
\bibitem [{\citenamefont {Huang}\ \emph {et~al.}(2019)\citenamefont {Huang},
  \citenamefont {Yang}, \citenamefont {Chen}, \citenamefont {Dzurak},\ and\
  \citenamefont {Goan}}]{Huang2019}%
  \BibitemOpen
  \bibfield  {author} {\bibinfo {author} {\bibfnamefont {C.-H.}\ \bibnamefont
  {Huang}}, \bibinfo {author} {\bibfnamefont {C.-H.}\ \bibnamefont {Yang}},
  \bibinfo {author} {\bibfnamefont {C.-C.}\ \bibnamefont {Chen}}, \bibinfo
  {author} {\bibfnamefont {A.~S.}\ \bibnamefont {Dzurak}},\ and\ \bibinfo
  {author} {\bibfnamefont {H.-S.}\ \bibnamefont {Goan}},\ }\bibfield  {title}
  {\bibinfo {title} {High-fidelity and robust two-qubit gates for quantum-dot
  spin qubits in silicon},\ }\href {https://doi.org/10.1103/PhysRevA.99.042310}
  {\bibfield  {journal} {\bibinfo  {journal} {Phys. Rev. A}\ }\textbf {\bibinfo
  {volume} {99}},\ \bibinfo {pages} {042310} (\bibinfo {year}
  {2019})}\BibitemShut {NoStop}%
\bibitem [{\citenamefont {Kuprov}\ and\ \citenamefont
  {Rodgers}(2009)}]{Kuprov2009}%
  \BibitemOpen
  \bibfield  {author} {\bibinfo {author} {\bibfnamefont {I.}~\bibnamefont
  {Kuprov}}\ and\ \bibinfo {author} {\bibfnamefont {C.~T.}\ \bibnamefont
  {Rodgers}},\ }\bibfield  {title} {\bibinfo {title} {Derivatives of spin
  dynamics simulations},\ }\href {https://doi.org/10.1063/1.3267086} {\bibfield
   {journal} {\bibinfo  {journal} {J. Chem. Phys.}\ }\textbf {\bibinfo {volume}
  {131}},\ \bibinfo {pages} {234108} (\bibinfo {year} {2009})}\BibitemShut
  {NoStop}%
\bibitem [{\citenamefont {Ball}\ \emph {et~al.}(2020)\citenamefont {Ball},
  \citenamefont {Biercuk}, \citenamefont {Carvalho}, \citenamefont {Chen},
  \citenamefont {Hush}, \citenamefont {Castro}, \citenamefont {Li},
  \citenamefont {Liebermann}, \citenamefont {Slatyer}, \citenamefont {Edmunds},
  \citenamefont {Frey}, \citenamefont {Hempel},\ and\ \citenamefont
  {Milne}}]{Ball2020}%
  \BibitemOpen
  \bibfield  {author} {\bibinfo {author} {\bibfnamefont {H.}~\bibnamefont
  {Ball}}, \bibinfo {author} {\bibfnamefont {M.~J.}\ \bibnamefont {Biercuk}},
  \bibinfo {author} {\bibfnamefont {A.}~\bibnamefont {Carvalho}}, \bibinfo
  {author} {\bibfnamefont {J.}~\bibnamefont {Chen}}, \bibinfo {author}
  {\bibfnamefont {M.}~\bibnamefont {Hush}}, \bibinfo {author} {\bibfnamefont
  {L.~A.~D.}\ \bibnamefont {Castro}}, \bibinfo {author} {\bibfnamefont
  {L.}~\bibnamefont {Li}}, \bibinfo {author} {\bibfnamefont {P.~J.}\
  \bibnamefont {Liebermann}}, \bibinfo {author} {\bibfnamefont {H.~J.}\
  \bibnamefont {Slatyer}}, \bibinfo {author} {\bibfnamefont {C.}~\bibnamefont
  {Edmunds}}, \bibinfo {author} {\bibfnamefont {V.}~\bibnamefont {Frey}},
  \bibinfo {author} {\bibfnamefont {C.}~\bibnamefont {Hempel}},\ and\ \bibinfo
  {author} {\bibfnamefont {A.}~\bibnamefont {Milne}},\ }\href@noop {} {\bibinfo
  {title} {Software tools for quantum control: Improving quantum computer
  performance through noise and error suppression}} (\bibinfo {year} {2020}),\
  \Eprint {https://arxiv.org/abs/2001.04060} {arXiv:2001.04060 [quant-ph]}
  \BibitemShut {NoStop}%
\bibitem [{\citenamefont {Cerfontaine}\ \emph {et~al.}(2014)\citenamefont
  {Cerfontaine}, \citenamefont {Botzem}, \citenamefont {DiVincenzo},\ and\
  \citenamefont {Bluhm}}]{Cerfontaine2014}%
  \BibitemOpen
  \bibfield  {author} {\bibinfo {author} {\bibfnamefont {P.}~\bibnamefont
  {Cerfontaine}}, \bibinfo {author} {\bibfnamefont {T.}~\bibnamefont {Botzem}},
  \bibinfo {author} {\bibfnamefont {D.~P.}\ \bibnamefont {DiVincenzo}},\ and\
  \bibinfo {author} {\bibfnamefont {H.}~\bibnamefont {Bluhm}},\ }\bibfield
  {title} {\bibinfo {title} {High-fidelity single-qubit gates for two-electron
  spin qubits in {G}a{A}s},\ }\href
  {https://doi.org/10.1103/PhysRevLett.113.150501} {\bibfield  {journal}
  {\bibinfo  {journal} {Phys. Rev. Lett.}\ }\textbf {\bibinfo {volume} {113}},\
  \bibinfo {pages} {150501} (\bibinfo {year} {2014})}\BibitemShut {NoStop}%
\bibitem [{\citenamefont {Haeberlen}\ and\ \citenamefont
  {Waugh}(1968)}]{Haeberlen1968}%
  \BibitemOpen
  \bibfield  {author} {\bibinfo {author} {\bibfnamefont {U.}~\bibnamefont
  {Haeberlen}}\ and\ \bibinfo {author} {\bibfnamefont {J.~S.}\ \bibnamefont
  {Waugh}},\ }\bibfield  {title} {\bibinfo {title} {Coherent averaging effects
  in magnetic resonance},\ }\href {https://doi.org/10.1103/PhysRev.175.453}
  {\bibfield  {journal} {\bibinfo  {journal} {Phys. Rev.}\ }\textbf {\bibinfo
  {volume} {175}},\ \bibinfo {pages} {453} (\bibinfo {year}
  {1968})}\BibitemShut {NoStop}%
\bibitem [{\citenamefont {Blanes}\ \emph {et~al.}(2009)\citenamefont {Blanes},
  \citenamefont {Casas}, \citenamefont {Oteo},\ and\ \citenamefont
  {Ros}}]{Blanes2009b}%
  \BibitemOpen
  \bibfield  {author} {\bibinfo {author} {\bibfnamefont {S.}~\bibnamefont
  {Blanes}}, \bibinfo {author} {\bibfnamefont {F.}~\bibnamefont {Casas}},
  \bibinfo {author} {\bibfnamefont {J.~A.}\ \bibnamefont {Oteo}},\ and\
  \bibinfo {author} {\bibfnamefont {J.}~\bibnamefont {Ros}},\ }\bibfield
  {title} {\bibinfo {title} {{The Magnus expansion and some of its
  applications}},\ }\href {https://doi.org/10.1016/j.physrep.2008.11.001}
  {\bibfield  {journal} {\bibinfo  {journal} {Phys. Rep.}\ }\textbf {\bibinfo
  {volume} {470}},\ \bibinfo {pages} {151} (\bibinfo {year}
  {2009})}\BibitemShut {NoStop}%
\bibitem [{\citenamefont {Magnus}(1954)}]{Magnus1954}%
  \BibitemOpen
  \bibfield  {author} {\bibinfo {author} {\bibfnamefont {W.}~\bibnamefont
  {Magnus}},\ }\bibfield  {title} {\bibinfo {title} {On the exponential
  solution of differential equations for a linear operator},\ }\href
  {https://doi.org/https://doi.org/10.1002/cpa.3160070404} {\bibfield
  {journal} {\bibinfo  {journal} {Commun. Pure. Appl. Math.}\ }\textbf
  {\bibinfo {volume} {7}},\ \bibinfo {pages} {649} (\bibinfo {year}
  {1954})}\BibitemShut {NoStop}%
\bibitem [{\citenamefont {Sza{\'{n}}kowski}\ \emph {et~al.}(2017)\citenamefont
  {Sza{\'{n}}kowski}, \citenamefont {Ramon}, \citenamefont {Krzywda},
  \citenamefont {Kwiatkowski},\ and\ \citenamefont
  {Cywi{\'{n}}ski}}]{Szakowski2017}%
  \BibitemOpen
  \bibfield  {author} {\bibinfo {author} {\bibfnamefont {P.}~\bibnamefont
  {Sza{\'{n}}kowski}}, \bibinfo {author} {\bibfnamefont {G.}~\bibnamefont
  {Ramon}}, \bibinfo {author} {\bibfnamefont {J.}~\bibnamefont {Krzywda}},
  \bibinfo {author} {\bibfnamefont {D.}~\bibnamefont {Kwiatkowski}},\ and\
  \bibinfo {author} {\bibfnamefont {{\L}.}~\bibnamefont {Cywi{\'{n}}ski}},\
  }\bibfield  {title} {\bibinfo {title} {Environmental noise spectroscopy with
  qubits subjected to dynamical decoupling},\ }\href
  {https://doi.org/10.1088/1361-648x/aa7648} {\bibfield  {journal} {\bibinfo
  {journal} {J. Phys. Condens. Matter}\ }\textbf {\bibinfo {volume} {29}},\
  \bibinfo {pages} {333001} (\bibinfo {year} {2017})}\BibitemShut {NoStop}%
\bibitem [{\citenamefont {Hangleiter}\ \emph
  {et~al.}(2021{\natexlab{b}})\citenamefont {Hangleiter}, \citenamefont {Le},\
  and\ \citenamefont {Teske}}]{Hangleiter2019a}%
  \BibitemOpen
  \bibfield  {author} {\bibinfo {author} {\bibfnamefont {T.}~\bibnamefont
  {Hangleiter}}, \bibinfo {author} {\bibfnamefont {I.~N.~M.}\ \bibnamefont
  {Le}},\ and\ \bibinfo {author} {\bibfnamefont {J.~D.}\ \bibnamefont
  {Teske}},\ }\href {https://doi.org/10.5281/zenodo.4575001} {\bibinfo {title}
  {{filter\_functions: A package for efficient numerical calculation of
  generalized filter functions to describe the effect of noise on quantum gate
  operations (version v1.0.0)}}},\ \bibinfo {howpublished}
  {\url{https://doi.org/10.5281/zenodo.4575001}} (\bibinfo {year}
  {2021}{\natexlab{b}})\BibitemShut {NoStop}%
\bibitem [{\citenamefont {Teske}(2021)}]{qopt2020}%
  \BibitemOpen
  \bibfield  {author} {\bibinfo {author} {\bibfnamefont {J.}~\bibnamefont
  {Teske}},\ }\href {https://github.com/qutech/qopt} {\bibinfo {title} {qopt: A
  simulation and quantum optimal control package}},\ \bibinfo {howpublished}
  {\url{https://github.com/qutech/qopt}} (\bibinfo {year} {2021})\BibitemShut
  {NoStop}%
\bibitem [{\citenamefont {Teske}\ \emph {et~al.}(2020)\citenamefont {Teske},
  \citenamefont {Cerfontaine},\ and\ \citenamefont {Bluhm}}]{Teske2020}%
  \BibitemOpen
  \bibfield  {author} {\bibinfo {author} {\bibfnamefont {J.}~\bibnamefont
  {Teske}}, \bibinfo {author} {\bibfnamefont {P.}~\bibnamefont {Cerfontaine}},\
  and\ \bibinfo {author} {\bibfnamefont {H.}~\bibnamefont {Bluhm}},\ }\bibfield
   {title} {\bibinfo {title} {qopt: A qubit simulation and quantum optimal
  control package}} (\bibinfo {year} {2020}),\ \bibinfo {note} {in
  preparation.}\BibitemShut {Stop}%
\bibitem [{\citenamefont {Cormen}\ \emph {et~al.}(2009)\citenamefont {Cormen},
  \citenamefont {Leiserson}, \citenamefont {Rivest},\ and\ \citenamefont
  {Stein}}]{Cormen2009}%
  \BibitemOpen
  \bibfield  {author} {\bibinfo {author} {\bibfnamefont {T.~H.}\ \bibnamefont
  {Cormen}}, \bibinfo {author} {\bibfnamefont {C.~E.}\ \bibnamefont
  {Leiserson}}, \bibinfo {author} {\bibfnamefont {R.~L.}\ \bibnamefont
  {Rivest}},\ and\ \bibinfo {author} {\bibfnamefont {C.}~\bibnamefont
  {Stein}},\ }\href@noop {} {\emph {\bibinfo {title} {{Introduction to
  Algorithms}}}},\ \bibinfo {edition} {3rd}\ ed.\ (\bibinfo  {publisher} {The
  MIT Press},\ \bibinfo {year} {2009})\BibitemShut {NoStop}%
\bibitem [{\citenamefont {Coppersmith}\ and\ \citenamefont
  {Winograd}(1990)}]{Coppersmith1990}%
  \BibitemOpen
  \bibfield  {author} {\bibinfo {author} {\bibfnamefont {D.}~\bibnamefont
  {Coppersmith}}\ and\ \bibinfo {author} {\bibfnamefont {S.}~\bibnamefont
  {Winograd}},\ }\bibfield  {title} {\bibinfo {title} {Matrix multiplication
  via arithmetic progressions},\ }\href
  {https://doi.org/https://doi.org/10.1016/S0747-7171(08)80013-2} {\bibfield
  {journal} {\bibinfo  {journal} {J. Symb. Comput.}\ }\textbf {\bibinfo
  {volume} {9}},\ \bibinfo {pages} {251} (\bibinfo {year} {1990})}\BibitemShut
  {NoStop}%
\bibitem [{\citenamefont {Zhu}\ \emph {et~al.}(1997)\citenamefont {Zhu},
  \citenamefont {Byrd}, \citenamefont {Lu},\ and\ \citenamefont
  {Nocedal}}]{Zhu1997}%
  \BibitemOpen
  \bibfield  {author} {\bibinfo {author} {\bibfnamefont {C.}~\bibnamefont
  {Zhu}}, \bibinfo {author} {\bibfnamefont {R.~H.}\ \bibnamefont {Byrd}},
  \bibinfo {author} {\bibfnamefont {P.}~\bibnamefont {Lu}},\ and\ \bibinfo
  {author} {\bibfnamefont {J.}~\bibnamefont {Nocedal}},\ }\bibfield  {title}
  {\bibinfo {title} {Algorithm 778: L-bfgs-b: Fortran subroutines for
  large-scale bound-constrained optimization},\ }\href
  {https://doi.org/10.1145/279232.279236} {\bibfield  {journal} {\bibinfo
  {journal} {ACM Trans. Math. Softw.}\ }\textbf {\bibinfo {volume} {23}},\
  \bibinfo {pages} {550–560} (\bibinfo {year} {1997})}\BibitemShut {NoStop}%
\bibitem [{\citenamefont {Morales}\ and\ \citenamefont
  {Nocedal}(2011)}]{Morales2011}%
  \BibitemOpen
  \bibfield  {author} {\bibinfo {author} {\bibfnamefont {J.~L.}\ \bibnamefont
  {Morales}}\ and\ \bibinfo {author} {\bibfnamefont {J.}~\bibnamefont
  {Nocedal}},\ }\bibfield  {title} {\bibinfo {title} {Remark on “algorithm
  778: L-bfgs-b: Fortran subroutines for large-scale bound constrained
  optimization”},\ }\href {https://doi.org/10.1145/2049662.2049669}
  {\bibfield  {journal} {\bibinfo  {journal} {ACM Trans. Math. Softw.}\
  }\textbf {\bibinfo {volume} {38}},\ \bibinfo {pages} {7} (\bibinfo {year}
  {2011})}\BibitemShut {NoStop}%
\bibitem [{\citenamefont {Nelder}\ and\ \citenamefont
  {Mead}(1965)}]{Nelder1965}%
  \BibitemOpen
  \bibfield  {author} {\bibinfo {author} {\bibfnamefont {J.~A.}\ \bibnamefont
  {Nelder}}\ and\ \bibinfo {author} {\bibfnamefont {R.}~\bibnamefont {Mead}},\
  }\bibfield  {title} {\bibinfo {title} {{A Simplex Method for Function
  Minimization}},\ }\href {https://doi.org/10.1093/comjnl/7.4.308} {\bibfield
  {journal} {\bibinfo  {journal} {Comput. J.}\ }\textbf {\bibinfo {volume}
  {7}},\ \bibinfo {pages} {308} (\bibinfo {year} {1965})}\BibitemShut {NoStop}%
\bibitem [{\citenamefont {Blaessle}()}]{constrNMPy}%
  \BibitemOpen
  \bibfield  {author} {\bibinfo {author} {\bibfnamefont {A.}~\bibnamefont
  {Blaessle}},\ }\href {https://github.com/alexblaessle/constrNMPy} {\bibinfo
  {title} {constrnmpy: Constrained nealder mead implemented in
  python}}\BibitemShut {NoStop}%
\bibitem [{\citenamefont {Pozza}\ \emph {et~al.}(2019)\citenamefont {Pozza},
  \citenamefont {Gherardini}, \citenamefont {Müller},\ and\ \citenamefont
  {Caruso}}]{Pozza2019}%
  \BibitemOpen
  \bibfield  {author} {\bibinfo {author} {\bibfnamefont {N.~D.}\ \bibnamefont
  {Pozza}}, \bibinfo {author} {\bibfnamefont {S.}~\bibnamefont {Gherardini}},
  \bibinfo {author} {\bibfnamefont {M.~M.}\ \bibnamefont {Müller}},\ and\
  \bibinfo {author} {\bibfnamefont {F.}~\bibnamefont {Caruso}},\ }\bibfield
  {title} {\bibinfo {title} {Role of the filter functions in noise
  spectroscopy},\ }\href {https://doi.org/10.1142/s0219749919410089} {\bibfield
   {journal} {\bibinfo  {journal} {Int. J. Quantum Inf.}\ }\textbf {\bibinfo
  {volume} {17}},\ \bibinfo {pages} {1941008} (\bibinfo {year}
  {2019})}\BibitemShut {NoStop}%
\bibitem [{\citenamefont {Kelly}\ \emph {et~al.}(2014)\citenamefont {Kelly},
  \citenamefont {Barends}, \citenamefont {Campbell}, \citenamefont {Chen},
  \citenamefont {Chen}, \citenamefont {Chiaro}, \citenamefont {Dunsworth},
  \citenamefont {Fowler}, \citenamefont {Hoi}, \citenamefont {Jeffrey},
  \citenamefont {Megrant}, \citenamefont {Mutus}, \citenamefont {Neill},
  \citenamefont {O'Malley}, \citenamefont {Quintana}, \citenamefont {Roushan},
  \citenamefont {Sank}, \citenamefont {Vainsencher}, \citenamefont {Wenner},
  \citenamefont {White}, \citenamefont {Cleland},\ and\ \citenamefont
  {Martinis}}]{Kelly2014}%
  \BibitemOpen
  \bibfield  {author} {\bibinfo {author} {\bibfnamefont {J.}~\bibnamefont
  {Kelly}}, \bibinfo {author} {\bibfnamefont {R.}~\bibnamefont {Barends}},
  \bibinfo {author} {\bibfnamefont {B.}~\bibnamefont {Campbell}}, \bibinfo
  {author} {\bibfnamefont {Y.}~\bibnamefont {Chen}}, \bibinfo {author}
  {\bibfnamefont {Z.}~\bibnamefont {Chen}}, \bibinfo {author} {\bibfnamefont
  {B.}~\bibnamefont {Chiaro}}, \bibinfo {author} {\bibfnamefont
  {A.}~\bibnamefont {Dunsworth}}, \bibinfo {author} {\bibfnamefont {A.~G.}\
  \bibnamefont {Fowler}}, \bibinfo {author} {\bibfnamefont {I.-C.}\
  \bibnamefont {Hoi}}, \bibinfo {author} {\bibfnamefont {E.}~\bibnamefont
  {Jeffrey}}, \bibinfo {author} {\bibfnamefont {A.}~\bibnamefont {Megrant}},
  \bibinfo {author} {\bibfnamefont {J.}~\bibnamefont {Mutus}}, \bibinfo
  {author} {\bibfnamefont {C.}~\bibnamefont {Neill}}, \bibinfo {author}
  {\bibfnamefont {P.~J.~J.}\ \bibnamefont {O'Malley}}, \bibinfo {author}
  {\bibfnamefont {C.}~\bibnamefont {Quintana}}, \bibinfo {author}
  {\bibfnamefont {P.}~\bibnamefont {Roushan}}, \bibinfo {author} {\bibfnamefont
  {D.}~\bibnamefont {Sank}}, \bibinfo {author} {\bibfnamefont {A.}~\bibnamefont
  {Vainsencher}}, \bibinfo {author} {\bibfnamefont {J.}~\bibnamefont {Wenner}},
  \bibinfo {author} {\bibfnamefont {T.~C.}\ \bibnamefont {White}}, \bibinfo
  {author} {\bibfnamefont {A.~N.}\ \bibnamefont {Cleland}},\ and\ \bibinfo
  {author} {\bibfnamefont {J.~M.}\ \bibnamefont {Martinis}},\ }\bibfield
  {title} {\bibinfo {title} {Optimal quantum control using randomized
  benchmarking},\ }\href {https://doi.org/10.1103/PhysRevLett.112.240504}
  {\bibfield  {journal} {\bibinfo  {journal} {Phys. Rev. Lett.}\ }\textbf
  {\bibinfo {volume} {112}},\ \bibinfo {pages} {240504} (\bibinfo {year}
  {2014})}\BibitemShut {NoStop}%
\bibitem [{\citenamefont {Cerfontaine}\ \emph {et~al.}(2020)\citenamefont
  {Cerfontaine}, \citenamefont {Botzem}, \citenamefont {Ritzmann},
  \citenamefont {Humpohl}, \citenamefont {Ludwig}, \citenamefont {Schuh},
  \citenamefont {Bougeard}, \citenamefont {Wieck},\ and\ \citenamefont
  {Bluhm}}]{Cerfontaine2020b}%
  \BibitemOpen
  \bibfield  {author} {\bibinfo {author} {\bibfnamefont {P.}~\bibnamefont
  {Cerfontaine}}, \bibinfo {author} {\bibfnamefont {T.}~\bibnamefont {Botzem}},
  \bibinfo {author} {\bibfnamefont {J.}~\bibnamefont {Ritzmann}}, \bibinfo
  {author} {\bibfnamefont {S.~S.}\ \bibnamefont {Humpohl}}, \bibinfo {author}
  {\bibfnamefont {A.}~\bibnamefont {Ludwig}}, \bibinfo {author} {\bibfnamefont
  {D.}~\bibnamefont {Schuh}}, \bibinfo {author} {\bibfnamefont
  {D.}~\bibnamefont {Bougeard}}, \bibinfo {author} {\bibfnamefont {A.~D.}\
  \bibnamefont {Wieck}},\ and\ \bibinfo {author} {\bibfnamefont
  {H.}~\bibnamefont {Bluhm}},\ }\bibfield  {title} {\bibinfo {title}
  {Closed-loop control of a {G}a{A}s-based singlet-triplet spin qubit with
  99.5{\%} gate fidelity and low leakage},\ }\href
  {https://doi.org/10.1038/s41467-020-17865-3} {\bibfield  {journal} {\bibinfo
  {journal} {Nature Communications}\ }\textbf {\bibinfo {volume} {11}},\
  \bibinfo {pages} {4144} (\bibinfo {year} {2020})}\BibitemShut {NoStop}%
\bibitem [{\citenamefont {Biercuk}\ \emph {et~al.}(2009)\citenamefont
  {Biercuk}, \citenamefont {Uys}, \citenamefont {VanDevender}, \citenamefont
  {Shiga}, \citenamefont {Itano},\ and\ \citenamefont
  {Bollinger}}]{Biercuk2009}%
  \BibitemOpen
  \bibfield  {author} {\bibinfo {author} {\bibfnamefont {M.~J.}\ \bibnamefont
  {Biercuk}}, \bibinfo {author} {\bibfnamefont {H.}~\bibnamefont {Uys}},
  \bibinfo {author} {\bibfnamefont {A.~P.}\ \bibnamefont {VanDevender}},
  \bibinfo {author} {\bibfnamefont {N.}~\bibnamefont {Shiga}}, \bibinfo
  {author} {\bibfnamefont {W.~M.}\ \bibnamefont {Itano}},\ and\ \bibinfo
  {author} {\bibfnamefont {J.~J.}\ \bibnamefont {Bollinger}},\ }\href@noop {}
  {\bibinfo {title} {High-fidelity quantum control using ion crystals in a
  penning trap}} (\bibinfo {year} {2009}),\ \Eprint
  {https://arxiv.org/abs/0906.0398} {arXiv:0906.0398 [quant-ph]} \BibitemShut
  {NoStop}%
\end{thebibliography}


%------------------------
%TC:endignore
\end{document}